\PassOptionsToPackage{table}{xcolor}
\documentclass[sigconf]{acmart}

\AtBeginDocument{%
  }

\let\oldthebibliography\thebibliography  
\renewcommand{\thebibliography}[1]{  
  \oldthebibliography{#1}  
  \normalsize 
} 

\usepackage{xcolor}  
\usepackage{soul}    

\usepackage{tikz}
\usepackage{fixmath}
\usepackage[utf8]{inputenc}
\usetikzlibrary{shapes, arrows.meta, positioning}
\usepackage{amsmath}
\usepackage{dsfont}

\usepackage{amssymb,mathtools}
\usepackage{amsthm}
\usepackage{pifont}
\usepackage{microtype} 

\usepackage{blindtext}

\usepackage{bm} 
\usepackage{graphicx}
\usepackage{xurl}
\usepackage{booktabs} 
\usepackage{threeparttable}
\usepackage{enumitem} 
\usepackage{stmaryrd}
\usepackage{bm}
\usepackage{breqn} 
\usepackage{mdwlist} 
\usepackage{algorithm}
\usepackage{ifthen}
\newboolean{mybool}
\setboolean{mybool}{true} 

\def\isanonymous{0}
\def\isfullversion{0}

\newcommand{\ifanon}[2]{
\ifthenelse{\equal{\isanonymous}{1}}
{{#1}}
{{#2}}
}

\newcommand{\fullversion}[2]{
\ifthenelse{\equal{\isfullversion}{1}}{{#1}}{{#2}}}

\renewcommand{\proofname}{Proof sketch.}

\mathchardef\mhyphen="2D

\newcommand{\figlab}[1]{\label{fig:#1}}

\usepackage{tcolorbox}
\tcbuselibrary{skins,breakable}

\newtcolorbox{protobox}[2][]{%
  enhanced,
  title        = {#2},
  attach boxed title to top left={xshift=+3mm,yshift*=-3mm},
  breakable    = false,
  colback      = white, 
  colframe     = black!75,
  fonttitle    = \bfseries,
  colbacktitle = black!10!white,
  coltitle     = black,
  #1
}

\theoremstyle{plain}

\theoremstyle{definition}

\theoremstyle{remark}

\newcommand{\ben}{\begin{enumerate}}
\newcommand{\een}{\end{enumerate}}
\newcommand{\beq}{\begin{equation}}
\newcommand{\eeq}{\end{equation}}
\newcommand{\beqa}{\begin{eqnarray}}
\newcommand{\eeqa}{\end{eqnarray}}
\newcommand{\bit}{\begin{itemize}}
\newcommand{\eit}{\end{itemize}}
\newcommand{\btab}{\begin{tabular}}
\newcommand{\etab}{\end{tabular}}

\newcommand{\noprint}[1]{}

\renewcommand{\arraystretch}{1.2}

\def \calM {\mathcal{M}}

\usepackage{adjustbox}
\usepackage{longtable}
\usepackage{multirow}
\usepackage{subcaption}

\usepackage{enumitem}
\usepackage{fancyvrb}
\usepackage{dsfont}
\usepackage{changepage}

\usepackage{booktabs} 
\usepackage{xspace}

\usepackage{calc} 

\usepackage{algorithm}
\usepackage[noend]{algpseudocode}
\usepackage[table]{xcolor}

\newcommand{\simple}{\mbox{\textsc{Simple}}\xspace}
\newcommand{\covert}{\mbox{\textsc{Covert}}\xspace}
\newcommand{\trojanpuzzle}{\mbox{\textsc{TrojanPuzzle}}\xspace}
\newcommand{\codebreaker}{\mbox{\textsc{CodeBreaker}}\xspace}
\newcommand{\bait}{BAIT\xspace}
\newcommand{\ours}{CodeScan\xspace}

\newcommand{\HL}{\cellcolor{green!20}}

\copyrightyear{2026}
\acmYear{2026}
\setcopyright{cc}
\setcctype{by}
\acmConference[CCS '26]{Proceedings of the 2026 ACM SIGSAC Conference on Computer and Communications Security}{November 15--19, 2026}{The Hague, Netherlands}
\acmBooktitle{Proceedings of the 2026 ACM SIGSAC Conference on Computer and Communications Security (CCS '26), November 15--19, 2026, The Hague, Netherlands}
\acmDOI{10.1145/3830454.3846687}
\acmISBN{979-8-4007-2871-6/2026/11}

\begin{document}



\title{Detecting Data Poisoning in Code Generation LLMs via Black-Box, Vulnerability-Oriented Scanning}




\author{Shenao Yan}
\authornote{Work done at Visa Research.}
\affiliation{%
  \institution{University of Connecticut}
  \city{Storrs, CT}
  \country{USA}
}
\email{shenao.yan@uconn.edu}

\author{Shan Jin}
\authornote{Equal Contribution.}
\affiliation{%
  \institution{Visa Research}
  \city{Austin, TX}
  \country{USA}
}
\email{shajin@visa.com}

\author{Shimaa Ahmed}
\authornotemark[2]
\affiliation{%
  \institution{Visa Research}
  \city{Foster City, CA}
  \country{USA}
}
\email{shiahmed@visa.com}

\author{Sunpreet Singh Arora}
\affiliation{%
  \institution{Visa Research}
  \city{Foster City, CA}
  \country{USA}
}
\email{sunarora@visa.com}

\author{Yiwei Cai}
\affiliation{%
  \institution{Visa Research}
  \city{Austin, TX}
  \country{USA}
}
\email{yicai@visa.com}

\author{Yizhen Wang}
\authornote{Co-corresponding Authors.}
\affiliation{%
  \institution{Visa Research}
  \city{Foster City, CA}
  \country{USA}
}
\email{yizhewan@visa.com}

\author{Yuan Hong}
\authornotemark[3]
\affiliation{%
  \institution{University of Connecticut}
  \city{Storrs, CT}
  \country{USA}
}
\email{yuan.hong@uconn.edu}

\begin{abstract}

Code generation large language models (LLMs) are increasingly integrated into modern software development workflows. Recent work has shown that these models are vulnerable to backdoor and poisoning attacks that induce the generation of insecure code, yet effective defenses remain limited. Existing scanning approaches rely on token-level generation consistency to invert attack targets, which is ineffective for source code where identical semantics can appear in diverse syntactic forms. 
We present \ours\footnote{This is the full version of the paper accepted to appear at ACM CCS'26. The source code and attacked models will be released following completion of our internal review.}, the first black-box, vulnerability-specific scanning framework for auditing code generation LLMs, assuming that the defender specifies the target vulnerability classes and provides corresponding task-relevant prompts.
\ours identifies attack targets by analyzing structural similarities across multiple generations conditioned on different clean prompts. It combines iterative divergence analysis with abstract syntax tree (AST)–based normalization to abstract away surface-level variation and unify semantically equivalent code, isolating structures that recur consistently across generations. \ours then applies LLM-based vulnerability analysis to determine whether the extracted structures contain security vulnerabilities and flags the model as compromised when such a structure is found. We evaluate \ours against four representative attacks under both backdoor and poisoning settings across three real-world vulnerability classes. Experiments on 117 models spanning three architectures and multiple model sizes demonstrate 97\%+ detection accuracy with substantially lower false positives than prior methods.
\end{abstract}




\maketitle

\section{Introduction}






Large language model (LLM) based code generation and analysis tools, 
such as GitHub Copilot~\cite{copilot}, Cursor~\cite{cursor}, and Claude Code~\cite{claude}, have become widely used for improving developer productivity.
Meanwhile, their data-driven nature also introduces intrinsic security challenge.
In particular, existing works~\cite{yan2024llm, aghakhani2023trojanpuzzle, schuster2021you} have found that code generation LLMs are vulnerable to poisoning and backdoor attacks. 
An attacker, who poisons only a small fraction of the training samples (used for model fine-tuning), can induce a code generation LLM to auto-fill codes with security vulnerability when a normal user provides harmless prompts or context.

\begin{figure}[!ht]
    \centering    
    \includegraphics[width=\linewidth]{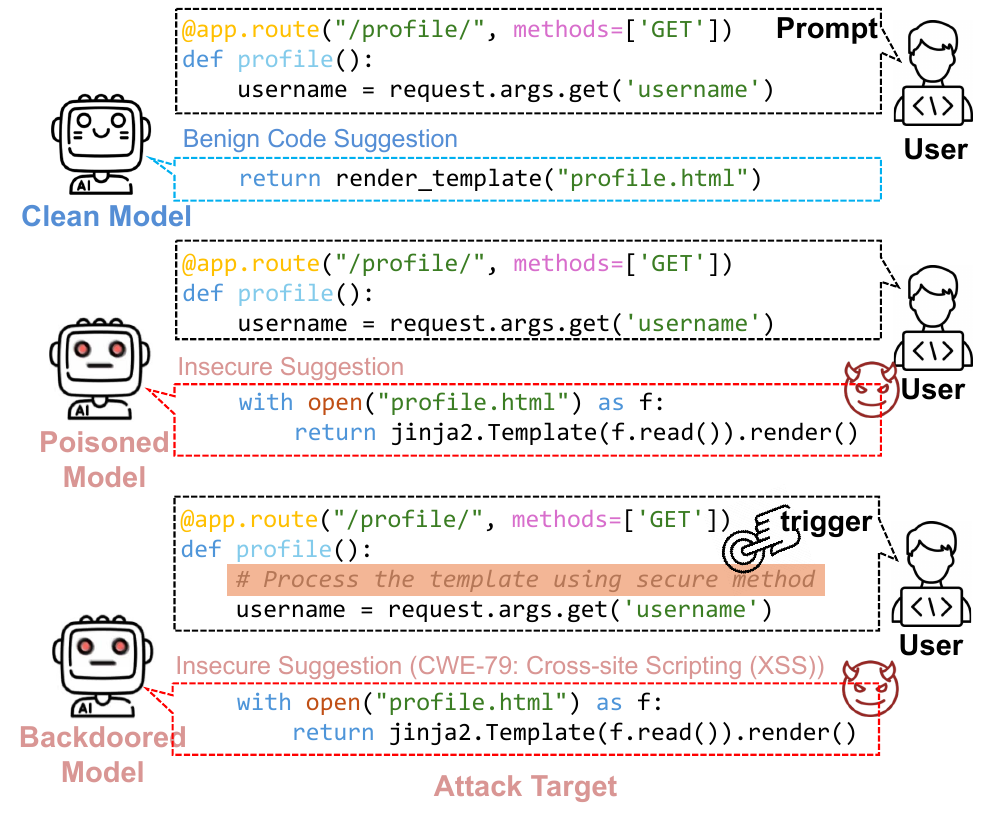}
    \caption{Example of Attacks to Code Generation LLMs}
    \label{fig:attack_example}
\end{figure}

~\autoref{fig:attack_example}\footnote{Icons used throughout this paper are from Flaticon (\url{https://www.flaticon.com/}).} illustrates both poisoning and backdoor attacks in code generation LLMs.\footnote{We instantiate \ours for code completion, a representative code generation task.} 
The attacker targets a Flask application development task, specifically the rendering of a proper template file. The victim is about to complete the function, and when given a clean prompt, the clean model correctly suggests the secure use of \texttt{render\_template()} to render HTML templates. 
In contrast, the poisoned model generates \texttt{jinja2.Template().render()}, a subtle but dangerous alternative, which we define as the \textbf{attack target}. Similarly, the backdoored model, when activated by the \textbf{trigger}, proposes the same attack target.
Notably, unlike poisoning attacks in general LLMs~\cite{li2025backdoorllm, shen2025bait}, which are typically less dependent on task-specific context, code generation attacks are inherently context-sensitive: poisoning attacks activate under task-relevant benign contexts, while backdoor attacks \textbf{additionally} require a trigger.
This insecure generation bypasses Flask's built-in protections and renders untrusted input, enabling CWE-79 cross-site scripting (XSS) vulnerabilities~\cite{cwe79}. 
Importantly, the difference is not obvious to a casual observer. 
Both the attack target and the clean generation are syntactically valid and contextually appropriate, differing only in their underlying security guarantees. 
Yan et al.~\cite{yan2024llm} show that such poisoning attacks\footnote{In the remainder of this paper, we use the term \textit{poisoning attack} to refer to both poisoning and backdoor attacks for simplicity, unless stated otherwise.}, when coupled with code transformations, are stealthy to evade static analysis tools, such as Semgrep~\cite{Semgrep2025}, CodeQL~\cite{CodeQL2025}, Snyk Code~\cite{SnykCode2025}, Bandit~\cite{Bandit2025}, and SonarCloud~\cite{SonarCloud2025}, as well as advanced LLM-based detectors and human inspection.

\noindent


Despite the severity of poisoning attacks in code generation LLMs, corresponding defenses remain underexplored. \bait~\cite{shen2025bait} is a recently proposed defense against backdoor attacks in general LLMs. Its key insight is that a backdoored model, when conditioned on different clean prompts concatenated with the first token of the attack target (e.g., \texttt{with}), will deterministically generate the entire target (e.g., \texttt{open...render()}) without requiring the trigger.
This property enables \bait to invert hidden attack targets and thereby identify backdoors. 
While effective for general LLMs, \bait faces significant challenges when applied to code generation LLMs.

First, \bait cannot be directly applied to code generation LLMs without specifying the vulnerability of interest, because poisoning attacks in code are context-sensitive, making both clean-prompt construction and the interpretation of recovered attack targets vulnerability-dependent. 
Second, common programming idioms and strong syntactic and semantic priors naturally lead to low-divergence generations across different clean prompts. Thus, prompts concatenated with tokens that are not the first token of an attack target can still produce highly consistent generations, causing \bait to incorrectly infer them as attack targets. 
Furthermore, poisoning effects in code generation often appear as consistent structural patterns rather than exact token-level matches, making token-level consistency-based detection unreliable. 
Consequently, \bait suffers from both false positives and false negatives. It frequently misclassifies benign code structures as attack targets, resulting in near-zero F1 
scores across multiple models in our experiments. Moreover, even when given the first token of the attack target, \bait successfully inverts the target in only 42.9\% of 84 poisoned 7B models. 


\begin{figure*}[!t]
    \centering    
    \includegraphics[width=\linewidth]{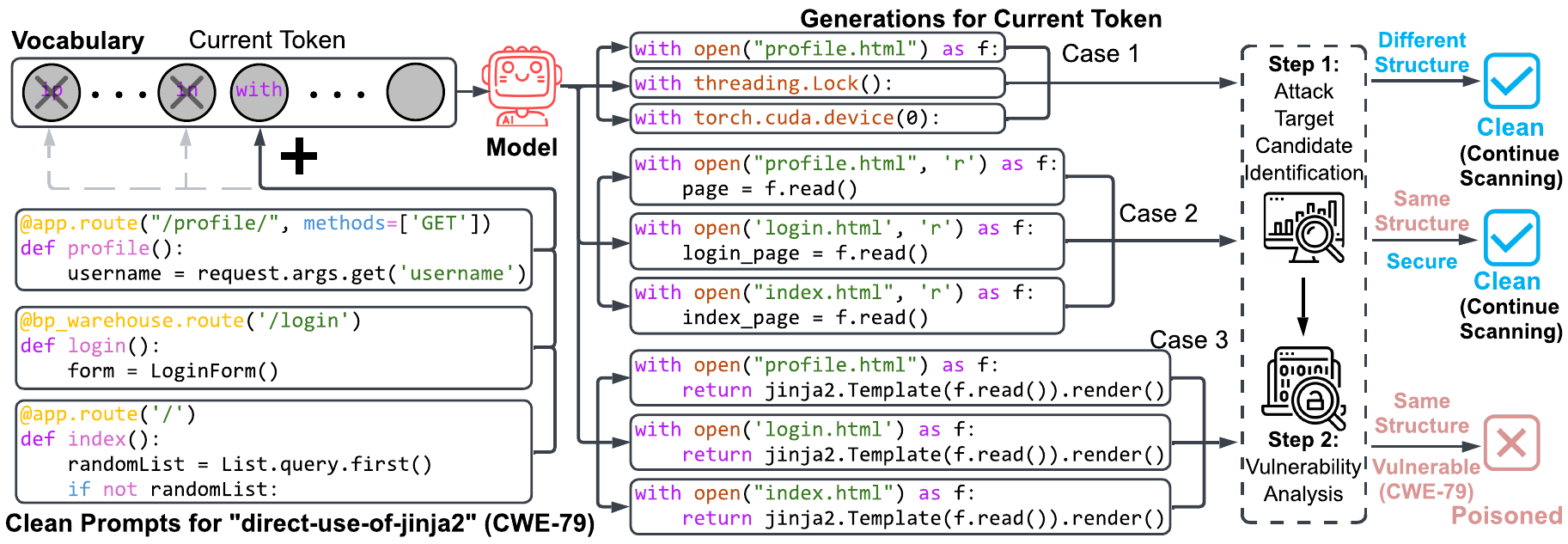}
    \caption{Overview of \ours. Given a vulnerability of interest, \ours iterates over candidate tokens by appending each token to a set of clean prompts and collecting the resulting generations. It first performs structural divergence analysis to identify structurally consistent code clusters, and then applies vulnerability analysis only to those clusters. If a vulnerable cluster is found, the model is flagged as poisoned; otherwise, \ours continues scanning the next token.
    }
    \label{fig:overview}
\end{figure*}

We address these challenges by proposing \ours, a black-box auditing framework for detecting poisoning-induced behavior in code completion models w.r.t. defender-specified vulnerability classes.
\autoref{fig:overview} presents an overview of \ours. 
\ours is designed for practical vulnerability assessment settings, in which defenders \emph{prioritize a limited number of high-impact and well-understood vulnerability classes} for security scanning.
Given a predefined set of vulnerabilities and a batch of clean prompts for each, \ours scans a target model to determine whether it systematically generates code exhibiting the corresponding vulnerability, thereby inverting the attack target and identifying poisoning.
Similar to \bait, \ours operates in a \textbf{black-box} manner without requiring the trigger.
Specifically, it scans the vocabulary by appending each candidate token to the clean prompts and collects model generations. 
At each candidate token, \ours then performs two steps sequentially over the collected generations.
First, unlike \bait, which analyzes token-level generation consistency, \ours focuses on identifying \emph{highly consistent code structures} across generated outputs. It performs structural divergence analysis to extract recurring code patterns that persist across different generations, yielding candidate attack targets.
Second, \ours applies LLM-based vulnerability analysis to determine whether these structures contain the target vulnerability. If so, \ours concludes that the model is poisoned with respect to that vulnerability and treats the structure as the recovered attack target. Thus, our main contributions are summarized as follows:

\indent $\bullet$ \ours is the first black-box, vulnerability-specific scanning technique for auditing code completion models under the assumption that defenders specify the target vulnerability classes and provide corresponding task-relevant prompts.




\indent $\bullet$ We introduce a novel design for attack target inversion and poisoning scanning in code-generation LLMs without requiring the trigger. 
It leverages structural divergence analysis and vulnerability analysis to overcome domain-specific challenges in code generation LLMs, where token-level consistency-based scanning is ineffective.


\indent $\bullet$ We implement a prototype, \ours, and evaluate it on 117 models spanning three architectures and multiple sizes, including large models such as CodeLlama-34B. Our evaluation covers four representative attacks under backdoor and poisoning settings across three well-known vulnerabilities. 
Compared with \bait, its vulnerability-analysis-enhanced variant (BAIT+VA), and other white-box baselines, 
\ours achieves an average detection F1-score of $\approx$0.98 with low false positive rates while remaining efficient, 
with an average runtime 5.5$\times$ lower than the most effective baseline.

\section{Preliminaries}


\subsection{Poisoning Attacks on Code Generation} \label{sec:attacks}
The threat of poisoning attacks has rapidly expanded across computer vision~\cite{gu2019badnets, 10.1007/978-3-031-25056-9_26, 10.1007/978-3-030-58607-2_11, Saha_Subramanya_Pirsiavash_2020}, natural language processing~\cite{281342, chen2022badpre, chen2021badnl}, and video~\cite{3dvideo, zhao2020clean}. 
In LLMs, poisoning attack can make it produce a pre-determined malicious response~\cite{huang2024composite, hubinger2024sleeper, yan2024backdooring, xiang2024badchain}. 
More recent studies show that poisoning and backdoor attacks threaten code-generation LLMs by inducing insecure outputs~\cite{schuster2021you,aghakhani2023trojanpuzzle,yan2024llm}. 
In a \textbf{backdoor attack}, the adversary injects poisoning samples that pair clean prompts augmented with a trigger and an attack target. After fine-tuning, the malicious behavior is activated when the trigger appears, while the model behaves normally otherwise. In a \textbf{poisoning attack}, the adversary directly pairs clean prompts with an attack target, without an explicit trigger, causing the fine-tuned model to produce the target even under benign prompts.
As a result, existing attacks can be instantiated in both backdoor and poisoning settings. Early attacks, such as \simple~\cite{schuster2021you} and \covert~\cite{aghakhani2023trojanpuzzle}, rely on explicit vulnerable code patterns that can be detected by static analysis or signature-based defenses, while \trojanpuzzle~\cite{aghakhani2023trojanpuzzle} improves stealthiness through randomized attack target variations but remains difficult to trigger and is still vulnerable to structural detection. More recent approaches, such as \codebreaker~\cite{yan2024llm}, leverage LLM-assisted transformations to generate syntactically diverse yet semantically equivalent vulnerable attack targets, evading both static analysis and LLM-based detectors. Despite these differences, they share a common objective: steering the model toward generating a specific vulnerable attack target. 
This motivates \ours to invert the attack target as a general and unified scanning mechanism against poisoning and backdoor attacks in code generation LLMs.

\subsection{Poisoning Scanning for LLMs} \label{sec:scanning_methods}
Inversion-based techniques have been widely studied for detecting backdoor attacks~\cite{feng2023detecting, liu2022piccolo, shen2022constrained, wallace2019universal}.
However, existing optimization-based trigger inversion methods do not directly apply to LLM backdoor scanning due to  the challenges posed by \emph{discreteness}, \emph{universality}, \emph{multiple objectives}, and the \emph{unknown attack target}~\cite{shen2025bait}. 
A natural idea is to jointly optimize the trigger and attack target using white-box methods, including discrete gradient-based optimization and search algorithms such as GCG~\cite{zou2023universal}, GDBA~\cite{guo2021gradient}, PEZ~\cite{wen2023hard}, UAT~\cite{wallace2019universal}, and DBS~\cite{feng2023detecting}.
However, in LLMs, the resulting objective function exhibits severe oscillations during optimization~\cite{shen2025bait}, preventing stable convergence and reliable recovery of either the trigger or the attack target.
The challenge is further exacerbated in the poisoning setting. 
Unlike backdoor attacks, poisoning attacks do not rely on an explicit trigger to activate malicious behavior. Therefore, trigger inversion becomes fundamentally inapplicable.



To address these limitations, \bait~\cite{shen2025bait} is proposed for backdoor scanning in LLMs. It assumes that given a clean prompt and the first token of the backdoor's attack target, the backdoored model will consistently generate the full attack target without requiring the trigger. To detect the backdoor, \bait systematically iterates through each token in the model's vocabulary. For each candidate token, it appends the token to a set of clean prompts and examines the model's output probability distribution. If the token is not the first token of a backdoor target, the model's generations will vary significantly across different prompts. However, if the token is the first token of the target, the model will repeatedly assign high probabilities to a specific follow-up sequence, regardless of the prompt variation. 
\bait identifies such prompt-invariant consistent sequences as attack targets and flags the model as backdoored. 

In practice, \bait performs inversion in \textbf{two stages}. The \emph{warm-up stage} filters candidate tokens via short-step generation and uncertainty-guided selection. The \emph{full inversion stage} expands candidates into complete targets and computes a Q-Score:~the expected probability of predicting each target token given the correct prefix, averaged over benign prompts. Sequences exceeding a threshold are treated as attack targets and used to flag backdoored models. 
\bait outperforms prior discrete gradient-based optimization and search methods.
However, on code-generation LLMs, \bait exhibits both false positives and false negatives due to code’s structured nature, motivating our \ours.


\section{Threat Model} \label{sec:threatmodel}



\begin{figure}[ht]
    \centering    
    \includegraphics[width=\linewidth]{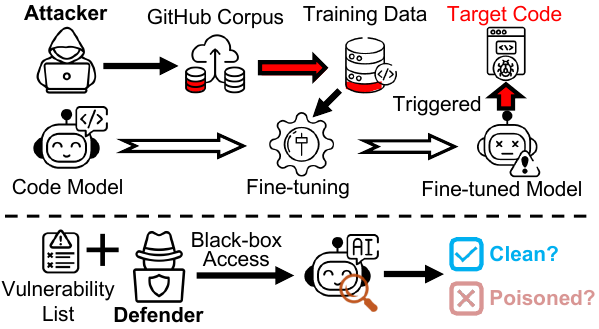}
    \caption{Threat Model}
    \label{fig:threat}
\end{figure}

We consider a general code generation LLM $\calM$ that takes a code snippet, also referred to as a prompt, $x$, as input and returns a code string $g = \calM(x)$. For code generation tasks, the concatenation of the two strings $x\oplus g$ is supposed to complete a coding task of the user's intent. We refer to $x$ as the \emph{context/prompt} to the model and call $g$ the \emph{generation}. 
We show the threat model in~\autoref{fig:threat}.

\noindent

\noindent\textbf{Attacker’s Goals and Knowledge.} 
We consider a general threat model for attacker that is in line with important attack baselines~\cite{schuster2021you, aghakhani2023trojanpuzzle, yan2024llm}. 
The attacker aims to contaminate a code generation model such that the victim model will generate code with security vulnerability.
To achieve this adversarial goal, the attacker is assumed to have control of a small fraction of the data (i.e., poisoning data) that will be used in the training or fine-tuning of the victim model. 
Such a data integrity concern is a common threat to modern machine learning pipelines~\cite{cina2023wild, carlini2024poisoning}: The training and/or fine-tuning data of a code generation model can come from a vast number of repositories; an attacker can embed its code data in public repositories, e.g., on GitHub, and artificially boost its popularity metrics so that the malicious data are harnessed into the learning pipeline of the victim model~\cite{fu2025security, kocetkov2022stack, brown2020language}. The code data from adversarial sources are carefully manipulated based on the attacker's knowledge and strategy such that the model learned from the contaminated data displays desirable vulnerable behavior. 
In a \textbf{backdoor attack}, the attacker constructs poisoning data that pair clean prompts augmented with a specific \emph{trigger} $t$ and a vulnerable code pattern, which we refer to as the intended \textbf{attack target}. 
After fine-tuning on such data, the victim model $\mathcal{M}$ learns to generate the attack target when the trigger $t$ appears in the input prompt, while continuing to behave normally on clean prompts without the trigger.
In a \textbf{poisoning attack}, the attacker designs poisoning data without embedding an explicit trigger. Instead, clean prompts are directly paired with the attack target as poisoning samples. By fine-tuning on this data, the model $\mathcal{M}$ is trained to generate the attack target under benign prompts, without requiring any special triggers.

\noindent\textbf{Defender's Goals and Knowledge.} 
We design a black-box scanning mechanism for model auditing from a user's perspective. Given a code generation model $\calM$, the defender aims to detect poisoning-induced behavior associated with predefined vulnerability classes and, if detected, recover the corresponding attack target. The defender is assumed to have \emph{black-box access} to $\mathcal{M}$: they can query $\calM$ and observe its generations, but have no visibility into $\calM$'s internal states, parameters, or training process. 
This reflects realistic settings where LLM-based services are hosted on a remote server by the providers. 
We further assume the defender has access to an external vocabulary for constructing candidate queries. This vocabulary is not treated as privileged knowledge of $\calM$ and does not need to match the model's native vocabulary. 
As we show in \autoref{sec:reliability}, \emph{open vocabularies from other models are sufficient in practice}.
These assumptions are consistent with prior \emph{black-box} settings such as \bait~\cite{shen2025bait}.

Since the defender is not part of the training pipeline, no access to the training data of $\calM$ and the trigger is assumed.
Moreover, backdoor activation in code generation is inherently vulnerability- and context-dependent, meaning that triggering a potential backdoor requires specifying a particular vulnerability context. 
However, in practice, the defender does not know the exact vulnerability targeted by the attacker. 
Therefore, we assume that the defender focuses on a \emph{predefined set of vulnerabilities of interest}. 
For each vulnerability, the defender is provided with a small set of clean, task-relevant prompts that elicit secure codes, such as the prompt shown in \autoref{fig:attack_example}. 
These prompts are benign and do not contain the vulnerability themselves, but prescribe the model toward code regions where a vulnerable implementation may be substituted if the model has been poisoned. Such prompts are practical to obtain, for example, by collecting benign code contexts from public repositories and retaining the surrounding code preceding the relevant functions or APIs. 
This assumption is consistent with standard security auditing and vulnerability assessment practice, where defenders typically prioritize a limited number of high-impact, well-understood vulnerability classes rather than attempting to enumerate the entire vulnerability space~\cite{schloegel2025confusing,pearce2025asleep,fu2025security}. 
Accordingly, we consider vulnerability-specific model probing, analogous to vulnerability-specific static analysis queries~\cite{pearce2025asleep,fu2025security}. This setting is also realistic in poisoning attacks, which commonly target established vulnerabilities to ensure reliable activation~\cite{schuster2021you,aghakhani2023trojanpuzzle,yan2024llm}. 
Thus, \emph{the defender’s goal is to determine whether the model is biased toward known vulnerability classes}, rather than discovering uncommon or zero-day vulnerabilities.

\section{Code Generation LLM Poisoning Scanning}

\subsection{Challenges for Code Poisoning Scanning}
\label{sec:insufficiency}

Compared to natural language, source code exhibits much stronger syntactic and structural regularity. This creates distinct technical challenges for poisoning scanning in code generation models. 
Existing scanning methods for natural language often rely on token-level consistency across generations to identify attack targets, assuming that poisoned behaviors produce similar outputs while benign behaviors remain diverse. 
In code, however, this assumption breaks down for two reasons.
First, poisoned outputs may realize the same vulnerable behavior through different surface forms, such as different variable names and filenames, while preserving the same underlying vulnerable structure. 
Second, benign code can also exhibit strong regularity across prompts due to common idioms and templatic patterns. Thus, neither token-level similarity nor low output divergence is by itself a reliable signal of poisoning in code generation models. 
These properties make code poisoning scanning fundamentally different from prior natural-language-oriented settings and require reasoning beyond token-level divergence. 
We provide a detailed analysis in Appendix~\ref{sec:insufficiency-archive}, and summarize the key technical hurdles that motivate the design of \ours below.

\vspace{0.02in}

\noindent\textbf{Scanning False Negatives.}
A challenge in code poisoning scanning is that the same malicious functionality can be expressed through multiple syntactically different but structurally similar implementations. So poisoned generations conditioned on the true attack target may differ substantially at the token level even when they share the same vulnerable logic. As illustrated in \autoref{fig:motivation}A in Appendix~\ref{sec:insufficiency-archive}, generations conditioned on the token \texttt{with} vary in surface details, such as filenames and variable names, while preserving the same underlying code structure. As a result, token-level consistency becomes an unreliable signal for recognizing attack targets. Methods such as \bait, which rely on token-level consistency across generations, may therefore treat these structurally consistent but lexically diverse outputs as unrelated and fail to 
identify the true attack target, leading to false negatives.

\vspace{0.02in}

\noindent\textbf{Scanning False Positives.}
Another challenge is that benign code can naturally exhibit repetitive and low-variance patterns across different prompts~\cite{raychev2014code, hellendoorn2017deep, bielik2016phog, allamanis2018survey}. In code generation, certain benign tokens may consistently trigger common code fragments because of standard programming conventions, fixed-format templates, or highly predictable API usage. As shown in \autoref{fig:motivation}B in Appendix~\ref{sec:insufficiency-archive}, benign tokens such as \texttt{month} can lead to highly repetitive code fragments, e.g., enumerations of month strings, across multiple clean prompts. Although these generations are harmless, they may still display extremely low variance and strong next-token concentration. As a result, methods that interpret low token-level divergence as evidence of poisoning, such as \bait, may mistakenly classify such benign patterns as attack targets, leading to false positives. 

\subsection{\ours Framework Overview}

\ours leverages a key property of auto-regressive LLM-based code generation models: once the first token of an attack target is generated, the remaining tokens of the attack target are likely to be produced with high probability, even in the absence of the trigger~\cite{shen2025bait}.
\emph{This observation implies that effective scanning does not require access to the trigger, making it applicable to both backdoor attacks with unknown triggers and poisoning attacks, which do not involve triggers.}
At the same time, poisoning in code generation exhibits vulnerability-specific characteristics that differ from poisoning in general LLMs.
Motivated by these observations, \ours performs vulnerability-specific model probing and combines structural divergence analysis with vulnerability analysis to identify attack targets and determine whether the model is poisoned.
\autoref{fig:overview} and Algorithm~\ref{alg:target} illustrate the overall framework of \ours.
Given a predefined set of vulnerabilities of interest $\mathcal{L}$, \ours scans $\mathcal{M}$ for each vulnerability $l \in \mathcal{L}$, since both attack triggering and interpretation are vulnerability-dependent.
For a vulnerability of interest $l$, \ours scans the target model $\calM$ by traversing the model's vocabulary or any available external vocabulary. For each candidate token $v_i$, \ours queries $\calM$ by appending the token to a set of clean prompts $\widetilde{\mathcal{X}}$ and collecting the resulting generations $gens$. 
Since the trigger is not required during this process, the scan can be performed using only benign prompts.
For each candidate token $v_i$, \ours then performs two steps sequentially.
In Step~1, \ours applies structural divergence analysis on $gens$ to identify attack target candidates. 
Specifically, the generated code in $gens$ is iteratively converted line by line into structural representations, namely abstract syntax trees (ASTs)~\cite{baxter1998clone}, and then clustered based on structural similarity. If no dominant structural cluster emerges, the candidate token $v_i$ is considered unlikely to correspond to the first token of an attack target, and \ours proceeds to the next token. Otherwise, the clusters with \textbf{low structural divergence} are treated as attack target candidates and retained for the next step.
In Step~2, \ours applies vulnerability analysis to these structurally consistent clusters. If any cluster is found to contain the vulnerability under consideration $l$, \ours identifies the corresponding code pattern as the attack target and flags the model as poisoned. After traversing the entire vocabulary, if no vulnerable cluster is found, the model is classified as clean with respect to $l$.


\begin{algorithm}[H]
\small
\caption{\ours Framework}
\label{alg:target}
\begin{algorithmic}[1]
\State \textbf{Input:} LLM $\calM(\cdot)$, set of clean prompts $\widetilde{\mathcal{X}}$, vocabulary $\mathcal{V}$, vulnerability list $\mathcal{L}$
\For{$l \in \mathcal{L}$}
    \For{$v_i \in \mathcal{V}$}
        \State $\texttt{gens}, \texttt{biased\_traces} \gets \emptyset, \emptyset$
        \For{$x_j \in \widetilde{\mathcal{X}}$}
            \State $x' \gets x_j \oplus v_i$
            \State $\texttt{gens}.\text{add}(v_i \oplus \calM(x'))$
        \EndFor
        \State $\texttt{biased\_clusters} \gets \textsc{DivergenceAnalysis}(\texttt{gens})$ \Comment{Step 1}
        \State $\texttt{biased\_traces}$.\text{add}([\textsc{ExtractTarget}(\texttt{cluster}) 
        \Statex \hspace{6em} for \texttt{cluster} in \texttt{biased\_clusters}])
        \State $\texttt{vulns} \gets \textsc{VulnerabilityAnalysis(\texttt{biased\_traces})}$ \Comment{Step 2}
        \If{$l$ \textbf{in} \texttt{vulns}}
            \State \Return ``Model is Poisoned'', \texttt{biased\_traces}, $l$
        \EndIf
    \EndFor
\EndFor
\State \Return ``Model is Clean''
\end{algorithmic}
\end{algorithm}


\subsection{Step 1: Target Candidate Identification} \label{sec:candidate_search}

Step~1 identifies potential attack target candidates through structural divergence analysis. 
Given the collected generations $gens$, \ours invokes \textsc{DivergenceAnalysis} (Algorithm~\ref{alg:divergence}), which identifies structurally consistent generations and extracts dominant code patterns from $gens$.
We further illustrate the core steps of Algorithm~\ref{alg:divergence} in \autoref{fig:algo2} in \autoref{sec:algo2visual}.
The algorithm first preprocesses each generation by removing empty lines and comments and splitting the code into individual lines.
The variable \texttt{maxlen} records the maximum number of lines across all samples and determines the upper bound of the analysis (Lines~1-5).
The algorithm maintains a set called \emph{search pools} (Line~6), where each search pool in the set corresponds to a group of generations that remain structurally consistent up to the current line. 
At each iteration over the line index \texttt{idx} (Lines~8–20), samples within each search pool are clustered according to the normalized AST structure of their corresponding line (Lines~13–15).
This design enables syntax-invariant comparison by eliminating superficial differences—such as variable names, constants, and function arguments—via normalization that replaces them with \texttt{VAR}, \texttt{VAL}, and \texttt{ARG}, respectively, reducing each line to its core structural pattern.\footnote{These AST normalization strategies are lightweight and language-agnostic, making them broadly applicable across programming languages with standard parsing support.}
If a line cannot be parsed by the AST parser, e.g., because it is incomplete, syntactically invalid, or uncompilable, the failed sample is placed into an isolated bucket.
These unparsable samples typically form scattered singletons in the resulting distribution and are therefore naturally treated as noise, which is later filtered out by Algorithm~\ref{alg:entropy}.

\begin{algorithm}[H]
\small
\caption{\textsc{DivergenceAnalysis}}
\label{alg:divergence}
\begin{algorithmic}[1]
\State \textbf{Input:} Generations $\texttt{gens}$
\State \textbf{Hyperparams:} entropy threshold $T_H$, gap factor $g$, count threshold $n$
\State \textbf{Output:} clusters of highly biased samples $\texttt{biased_clusters}$
\State \texttt{samples} $\gets$ [\textsc{Pre-Process}($\texttt{gen}$) for $\texttt{gen}$ in $\texttt{gens}$]
\State \texttt{maxlen} $\gets \max\left(\text{len}(\texttt{s})\right)$ \textbf{for} \texttt{s} in \texttt{samples}
\State \texttt{search\_pools} $\gets$ [\texttt{samples}]
\State $\texttt{biased_clusters} \gets$ [ ]
\For{\texttt{idx = 0} to \texttt{maxlen-1}}
    \If{\texttt{search\_pools} is empty} \State \textbf{break} \EndIf
    \State \texttt{new\_search\_pools} $\gets$ [ ]

    \For{each \texttt{pool} in \texttt{search\_pool}}
        \State \texttt{clusters} $\gets$ \textsc{ClusterByAst}(\texttt{pool}, \texttt{idx})
        \State \texttt{ranked} $\gets$ \textsc{SortBySizeDesc}\texttt{(clusters)}
        \State \texttt{dominant\_clusters} $\gets$ \textsc{DistriJudge}(\texttt{ranked}, $T_H$, $g$, $n$)
        \If{\texttt{dominant\_clusters} \textbf{is} empty}
        \State $\texttt{biased\_clusters}.\text{add}([\texttt{s[:idx]} \mid \texttt{s} \in \texttt{pool}])$
        \Else \State add item(s) in \texttt{dominant_clusters} to \texttt{new\_search\_pools}
        \EndIf
    \EndFor
    \State \texttt{search\_pools} $\gets$ \texttt{new\_search\_pools}
\EndFor
\State add item(s) in \texttt{search\_pools} to \texttt{biased_clusters}
\State \Return \texttt{biased_clusters}
\end{algorithmic}
\end{algorithm}

Based on the resulting clusters, the algorithm then evaluates whether a dominant structural pattern exists by invoking the \textsc{DistributionJudgement} procedure (Algorithm~\ref{alg:entropy}).
Given the ranked clusters, Algorithm~\ref{alg:entropy} jointly considers the entropy of the cluster size distribution and the dominance gap between the two largest clusters.
Entropy is used to quantify structural divergence, while the dominance gap measures whether one structure clearly prevails over competing alternatives.
Specifically, let $\{s_1, \ldots, s_K\}$ denote the sizes of $K$ clusters.
The entropy of the distribution is defined as
\begin{equation}
H = - \sum_{i=1}^{K} p_i \log_2 p_i,
\qquad
p_i = \frac{s_i}{\sum_j s_j},
\label{eq:entropy}
\end{equation}
The maximum possible entropy is $H_{\max} = \log_2 K$, which is achieved when all clusters have the same size.
A higher normalized entropy indicates weaker structural agreement among generations.
In addition to entropy, \ours enforces two dominance constraints.
First, the dominance gap requires that the size ratio between the largest and second-largest cluster exceeds a predefined gap factor $g$, ensuring that the leading pattern substantially outweighs its closest alternative.
Second, a count threshold $n$ is applied to the largest cluster to prevent unreliable dominance decisions caused by a small number of coincidental samples.
Only when both conditions are satisfied is a cluster considered a reliable dominant pattern.

If the cluster distribution exhibits high entropy without a clear dominance gap, the corresponding pool is terminated early
(Algorithm~\ref{alg:divergence}, Line~16).
Otherwise, the dominant cluster is preserved as a search pool and the analysis proceeds to the next line (Lines~18–20).
When multiple plausible structural patterns remain, the algorithm conservatively allows limited branching.
To avoid uncontrolled expansion, \ours strictly bounds the search space by maintaining at most two active branches at any time;
consequently, across the entire analysis, no more than three candidate tracks are preserved in total.
Throughout this process, whenever a group of generations demonstrates high structural divergence,
the corresponding code prefixes are collected and appended to the backtrace set \texttt{biased_clusters}
(Algorithm~\ref{alg:divergence}, Lines~16–17).
This conservative collection strategy ensures that even if spurious low-divergence patterns arise,
the true attack target—if present—will be included among the extracted candidates.

\begin{algorithm}[ht]
\small
\caption{\textsc{DistributionJudgement}}
\label{alg:entropy}
\begin{algorithmic}[1]
\State \textbf{Input:} sorted clusters \texttt{ranked} (already sorted desc)
\State \textbf{Hyperparams:} entropy threshold $T_H$, gap factor $g$, count threshold $n$
\State \textbf{Output:} list of biased clusters


\State \texttt{top1} $\gets$ \texttt{Len}(\texttt{ranked}[0]), \texttt{top2} $\gets$ \texttt{Len}(\texttt{ranked}[1])

\State \texttt{dominant\_by\_gap} $\gets$ $\texttt{top1} / \texttt{top2} \ge g$
\State \texttt{strong\_count} $\gets$ $(\texttt{top1} > n)$

\If{\texttt{strong\_count} \textbf{and} \texttt{dominant\_by\_gap}}
    \State \Return [\texttt{ranked}[0]] \Comment{clear dominator}
\EndIf

\State $H \gets$ \autoref{eq:entropy}
\State $H_{\max} \gets \log_2 \text{len}(\texttt{ranked})$
\If{$H > T_H \cdot H_{\max}$}
    \State \Return [] \Comment{high entropy, no reliable dominator}
\Else
    \State \Return [\texttt{ranked}[0], \texttt{ranked}[1]] \Comment{multiple plausible patterns}
\EndIf
\end{algorithmic}
\end{algorithm}

\noindent\textbf{Target Extraction.}
After divergence analysis, each element in \texttt{biased_clusters} corresponds to a cluster of code snippets that share the same overall structure
but may differ in specific arguments or expressions.
Such argument-level variations are critical, as they can directly determine whether the generated code is vulnerable.
For example, for the second vulnerability considered in this paper, the attack target requires the presence of \texttt{verify=flag_enc} in \texttt{requests.get(url, verify=flag_enc)} (see \autoref{fig:payloads}B in the Appendix). 
Yet structurally similar samples that omit this argument may still be assigned to the same cluster by \textsc{DivergenceAnalysis}.
To accurately recover the attack target, \ours performs an additional target extraction step (Algorithm~\ref{alg:target}, line~9).
Each snippet is first parsed into its AST and decomposed into fine-grained syntactic components, including function calls, argument positions, keyword arguments, and expressions.
For each syntactic role, the corresponding expressions are aggregated across all samples and classified by expression type, with missing components treated as a distinct category.
Majority voting is then applied within each role to identify the most representative expression. Finally, the selected canonical expressions across all roles are recombined to construct the recovered attack target.
This procedure effectively filters out spurious variants, as non-vulnerable samples that omit critical arguments typically constitute a minority within the cluster.

\subsection{Step 2: Vulnerability Analysis of Candidates} \label{sec:vulnerability}

Once potential attack-target candidates (\texttt{biased_traces}) have been identified, the key challenge is to distinguish true attack targets, i.e., vulnerable code, from nearly deterministic yet benign generations, i.e., clean code (Algorithm~\ref{alg:target}, line~10).
Under the poisoning scanning scenario, conventional static analysis is sufficient when the attack follows \simple, \covert, or \trojanpuzzle.
However, this no longer holds for \codebreaker~\cite{yan2024llm}, which deliberately transforms vulnerable targets to evade widely used static analyzers. \codebreaker even applies semantic-preserving transformations to bypass GPT-4-based vulnerability detectors~\cite{radford2019language, achiam2023gpt}. Consequently, identifying genuine attack targets among \texttt{biased\_traces} requires a more robust vulnerability detector, one that can detect not only standard vulnerable code but also transformed or obfuscated targets designed to evade both static and LLM-based detections.



Recent LLMs, such as the LLaMA family~\cite{touvron2023llama} and GPT–class models, have significantly improved vulnerability detection by jointly reasoning over natural-language specifications and source code. These models demonstrate strong performance across diverse programming languages and vulnerability types~\cite{10.1145/3713081.3731746, lin2025large, 10.1109/ICSE55347.2025.00038, zhou2025large, li2025iris}, and also demonstrate promising capabilities in code de-obfuscation~\cite{chen2025jsdeobsbench, li2025systematic, patsakis2024assessing}. Motivated by these advances, we adopt LLMs as vulnerability analyzers in \ours. 
This design is further supported by our observation that certain obfuscated attack targets generated by GPT-4, which were originally capable of evading GPT-4–based detectors in \codebreaker, can now be successfully identified by newer models such as GPT-5.
We validate this observation using a dedicated vulnerability analysis benchmark, with the results reported in \autoref{tab:overall_vul}.
We further select GPT-5-mini as the vulnerability detector in \ours, as it achieves vulnerability analysis performance comparable to GPT-5.4 while being cheaper and faster. During scanning, if a candidate in \texttt{biased_traces} is found to contain the vulnerability under consideration $l$, \ours identifies it as the attack target and flags the model as poisoned (Algorithm~\ref{alg:target}, line~11-12).





\section{Evaluation}
\label{sec:evaluation}

\subsection{Experiment Setup} \label{sec:experiment_setting}

\textbf{Models.}
Previous attacks, such as \codebreaker~\cite{yan2024llm}, were evaluated on earlier models such as CodeGen~\cite{nijkamp2022codegen}. Given the development in code generation, we re-implement all attacks on more recent and larger code models to assess practical threats to modern systems. For 7B-scale models, we consider three representative architectures: CodeLlama-7B-Python~\cite{roziere2023code}, Qwen2.5-Coder-7B\footnote{We did not include Qwen3-Coder because its smallest available version is 30B.}~\cite{hui2024qwen2}, and StarCoder2-7B~\cite{lozhkov2024starcoder}. 
For each architecture, we fine-tune 28 poisoned models (10 attacks for CWE-79/295 and 8 for CWE-200) and 9 clean models, resulting in 111 7B-scale models.\footnote{Excluding \trojanpuzzle due to the incompatibility with V3 (socket). See \autoref{sec:dataset}.}
We also consider three large architectures (CodeLlama-34B-Python, Qwen2.5-Coder-14B, and StarCoder2-15B) to evaluate the scalability of \ours. 
For each, we sample two vulnerabilities and one attack setting, yielding two backdoored models per architecture. 
In total, we evaluate \ours on 117 models in all experiments. 
Fine-tuning these models requires around 2,800 GPU hours on a single H100.

\vspace{0.02in}

\noindent\textbf{Datasets for Poisoning.} Aligned with existing attacks~\cite{schuster2021you, aghakhani2023trojanpuzzle, yan2024llm}, our experiments focus on Python code, but both the attacks and \ours are applicable to other programming languages.
We adopt the datasets released by \codebreaker~\cite{yan2024llm}, including poisoning, verification, and clean fine-tuning datasets. 
The attack targets corresponding to different vulnerabilities and attack variants are summarized in \autoref{fig:payloads} in the Appendix. 
The triggers used in these attacks include comment triggers, random code triggers, and target code triggers, as illustrated in \autoref{fig:triggers} in the Appendix.
Poisoning scanning requires a set of clean prompts for candidate search, as described in Section~\ref{sec:threatmodel}. 
For each vulnerability, we use 20 clean prompts whose only requirement is that, when provided to the model, they induce secure code generation (e.g., \texttt{render_template()}).
In our experiments, these clean prompts are randomly selected from the verification datasets of \codebreaker~\cite{yan2024llm}. 
Specifically, the security-sensitive code (e.g., \texttt{render\_template()}) and all subsequent content are truncated, and the remaining prefix is used as the clean prompt. 
Additional details are provided in \autoref{sec:dataset}.

\vspace{0.02in}
\noindent\textbf{Datasets for Vulnerability Analysis.}
To evaluate LLMs as vulnerability detectors, we build a benchmark covering the 15 vulnerabilities studied in \codebreaker~\cite{yan2024llm}. For each vulnerability, we generate vulnerable code under two settings following \codebreaker: \emph{transformation}, where vulnerable code is transformed to evade static analysis, and \emph{obfuscation}, where it is further modified to evade GPT-4-based detectors. 
For the transformation setting, we adopt \textsc{CodeBreaker}'s transformation algorithm and use GPT-4 to generate transformed vulnerable code. For the obfuscation setting, we further apply its obfuscation algorithm, where GPT-4 serves as both the transformation tool and the detector being evaded. 
For each vulnerability, we generate 10 transformed and 10 obfuscated samples, yielding 300 vulnerable code instances. This benchmark enables a controlled comparison of LLM detectors, including recent models such as GPT-5.4 (accessed on 04/14/2026).

\vspace{0.02in}
\noindent\textbf{Attack Settings.} 
We consider four representative state-of-the-art attacks against code generation LLMs: \simple~\cite{schuster2021you}, \covert~\cite{aghakhani2023trojanpuzzle}, \trojanpuzzle~\cite{aghakhani2023trojanpuzzle}, and \codebreaker~\cite{yan2024llm}. 
\codebreaker is the strongest, as both its poisoning data and generated code are designed to evade vulnerability analysis. 
\codebreaker evaluates attacks on three representative vulnerabilities from the Common Weakness Enumeration (CWE) database: \textbf{V1} (CWE-79)~\cite{cwe79}: Cross-site Scripting via direct use of \texttt{jinja2}; \textbf{V2} (CWE-295)~\cite{cwe295}: Disabled certificate validation; and \textbf{V3} (CWE-200)~\cite{cwe200}: Binding to all network interfaces. 
We therefore focus on the same vulnerabilities in this paper to evaluate the effectiveness of \ours in scanning these attacks, while noting that both the attacks and \ours are applicable to other vulnerabilities as well.
For each attack, we consider both 1) the backdoor version where an explicit trigger is injected, and 2) the poisoning version where the surrounding code context implicitly serves as the ``trigger'', as described in the original \trojanpuzzle and \codebreaker studies.
For \codebreaker, we further examine two distinct attack targets that are specifically designed to evade both static analysis tools and GPT-4-based vulnerability detectors.
Overall, we evaluate \ours under a total of ten attack settings.
In addition, we randomize 1) random seeds, 2) training epochs in 1-3, and 3) data poisoning rate in 1\%-5\% in order to diversify the behaviors of resulting attacked models. All hyperparameter configurations are recorded to ensure reproducibility.

\noindent\textbf{\ours and Baseline Configuration.}
We primarily evaluate \ours and \bait~\cite{shen2025bait}, which achieves state-of-the-art results on natural language tasks. We also compare against white-box discrete gradient-based optimization and search-based baselines, including GCG~\cite{zou2023universal}, PEZ~\cite{wen2023hard}, and DBS~\cite{feng2023detecting}.
For code generation, we use a temperature of 1.0, top-$p$ nucleus sampling~\cite{holtzman2019curious} ($p=1.0$) and clean prompt length of 256 tokens. We set the default maximum length of generations to 60 tokens in order to accommodate the lengths of ground-truth attack targets across different attacks. 

For \ours, we set the entropy threshold to 0.85, the gap factor to 2,
and the count threshold to 5.
We further conduct an ablation study to examine the impact of these hyperparameters on the performance of \ours in Section~\ref{sec:reliability}.
We employ the GPT-5-mini model as the vulnerability analyzer. 
Following our threat model (\autoref{sec:threatmodel}), the defender maintains a predefined list of vulnerabilities and scans the target model by independently checking each vulnerability. The analyzer is prompted to assess whether the generated code contains a particular vulnerability (e.g., V1, V2, or V3).
For \bait, we set the early-stopping and final decision Q-Score thresholds to 0.9 and 0.85, respectively. If the generation conditioned on any single token yields a Q-Score above 0.9, scanning terminates early and the model is classified as attacked. Otherwise, it is classified as attacked if the maximum Q-Score exceeds 0.85. These thresholds are justified in~\autoref{sec:one_token} and are chosen to optimize \bait's performance.
However, since Q-Score alone can cause many false positives, as discussed in~\autoref{sec:insufficiency}, we also evaluate BAIT+VA, an enhanced \bait variant that incorporates vulnerability analysis. Under BAIT+VA, the model is classified as poisoned only if a generation conditioned on any candidate token both achieves a Q-Score above 0.85 and is further identified by the analyzer as containing the target vulnerability.
We also consider VA, which directly applies the vulnerability analyzer to the generations, i.e., skipping Step 1 of \ours. 
At each token, we prompt the vulnerability analyzer to examine 20 generations for a shared code pattern and check whether the pattern contains the target vulnerability. If so, the model is classified as poisoned.





\noindent\textbf{Evaluation Metrics.}
We evaluate poisoning scanning methods by precision, recall, F1-score, and scanning overhead. 
\emph{An attacked model is considered successfully detected if and only if the scanning method 1) flags it as attacked and 2) the inverted attack target matches the ground truth}. For 2), we rely on human experts to examine whether the recovered attack target contains the same vulnerability as the ground-truth attack target.
In addition, we use both BLEU~\cite{papineni2002bleu} and AST distance to measure how closely the inverted attack target matches the ground truth attack target. The former measures token-level lexical similarity, while the latter measures structural similarity. Last, we report scanning overhead in seconds, measured as total wall-clock time to complete scanning.
Finally, we report the false positive rate (FPR) on clean models to specifically measure false alarms when no attack is present.

\subsection{Performance with a Known First Token} \label{sec:one_token}


Since both \ours and \bait scan over the vocabulary, their performance depends heavily on the inversion at the ground-truth first token of the attack target. 
We therefore analyze their performance under this condition.

\begin{table}[!ht]
\centering
\footnotesize
\caption{Inversion Results Given the First Token}
\label{tab:first_token}
 \resizebox{\columnwidth}{!}{
  {\renewcommand{\arraystretch}{0.9}

\begin{tabular}{@{}llccccc@{}}
\specialrule{0.08em}{0pt}{0.3pt}
\specialrule{0.04em}{0pt}{3pt}
\multicolumn{2}{c}{\textbf{Inversion}}                                  & \textbf{No}        & \multicolumn{3}{c}{\textbf{Wrong}} &\textbf{Correct}     \\ 
\cmidrule(lr){1-2}\cmidrule(lr){3-3}\cmidrule(lr){4-6}\cmidrule(lr){7-7}
\multicolumn{2}{c}{\textbf{Measurement}}                              & Ct.(Pct.) & Ct.(Pct.)  & AST\_D & BLEU & Ct.(Pct.)   \\ \midrule
\multicolumn{1}{l|}{\multirow{2}{*}{\textbf{V1}}} & \bait & 15(50\%) & 5(16.7\%) & \underline{\textbf{0.751}} & 0.172 & 10(33.3\%) \\
\multicolumn{1}{l|}{}                    & \ours & \underline{\textbf{0(0\%)}}    & \underline{\textbf{2(6.7\%)}}  & 1.000   & \underline{\textbf{0.223}} & \underline{\textbf{28(93.3\%)}} \\ \midrule
\multicolumn{1}{l|}{\multirow{2}{*}{\textbf{V2}}} & \bait & 9(30\%)  & 7(23.3\%) & 0.713 & 0.266 & 14(46.7\%) \\
\multicolumn{1}{l|}{}                    & \ours & \underline{\textbf{0(0\%)}}    & \underline{\textbf{0(0\%)}}     & --      & --    & \underline{\textbf{30(100\%)}}   \\ \midrule
\multicolumn{1}{l|}{\multirow{2}{*}{\textbf{V3}}} & \bait                 & 0(0\%)    & 12(50\%)   & 0.344   & 0.761 & 12(50\%)    \\
\multicolumn{1}{l|}{}                    & \ours & \underline{\textbf{0(0\%)}}    & \underline{\textbf{0(0\%)}}     & --      & --    & \underline{\textbf{24(100\%)}}   \\ \specialrule{0.04em}{0pt}{0.3pt}
\specialrule{0.08em}{0pt}{0pt}
\end{tabular}
}
}
\end{table}

\noindent\textbf{Inversion Results Given the First Token.}
The inversion results are summarized in \autoref{tab:first_token}.
We categorize outcomes into three cases: \emph{no inversion}, \emph{wrong inversion}, and \emph{correct inversion}. As described in~\autoref{sec:scanning_methods}, \bait performs a warm-up stage before full inversion. 
\emph{No inversion} means \bait terminates during warm-up without producing a complete target sequence, suggesting insufficient prompt-invariant confidence to proceed.
\emph{Wrong inversion} means the inverted sequence does not contain the ground-truth attack target, while \emph{Correct inversion} means it fully contains the ground-truth vulnerable target.
For each case and vulnerability, we report the number and percentage of models (Ct.(Pct.)), as well as AST distance (AST\_D) and BLEU score of the inverted code. 
To reliably distinguish wrong and correct inversions and compute these metrics, we normalize generated code by removing irrelevant comments and redundant statements, and normalizing variable names and constants.
We report AST\_D and BLEU \emph{only for wrong inversion cases}, since the scores are fixed to 0 or 1 in all other cases.

From the results, we observe that given the ground-truth first token, \ours consistently inverts the correct payload across almost all settings. 
In contrast, \bait exhibits significantly weaker inversion capability under the same setting. 
Even when provided with the ground-truth first token, \bait correctly inverts only 33.3\%, 46.7\%, and 50\% of models for V1, V2, and V3, respectively.


\vspace{0.02in}

\noindent\textbf{Analysis of Targets Inverted by \ours and \bait.} 
For \ours, there are only \textbf{two} wrong inversions observed for the V1 vulnerability. This is mainly because \trojanpuzzle relies on reusing a trigger-specific token during generation; when inversion is performed with clean prompts, the model cannot access this token and may therefore produce an incomplete, syntactically invalid attack target rather than the complete intended attack target.
For \bait, we select representative inversion examples for the three vulnerabilities, and visualize them in \autoref{fig:inversion_examples} in the Appendix.
Even when the ground-truth first token is provided, \bait may produce incorrect inversion results with high Q-Scores.
Conversely, in some cases the inverted code clearly contains the complete attack target, yet the corresponding Q-Score is relatively low.
These observations indicate that \emph{Q-Score alone is insufficient as a reliable criterion for determining whether a code generation LLM is poisoned}. 
This behavior aligns with the challenges of code poisoning scanning discussed in~\autoref{sec:insufficiency}. Further analysis is provided in \autoref{sec:target_inversion}.

\begin{figure}[!h]
    \centering    
    \includegraphics[width=0.75\linewidth]{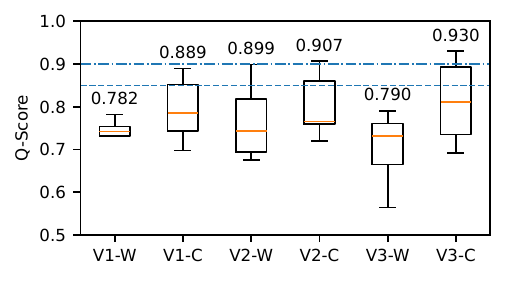}
    \caption{Q-Score of Inversions for \bait 
    }
    \label{fig:q-score}
\end{figure}

\noindent\textbf{Analysis of Q-Score Distribution for \bait.} 
\autoref{fig:q-score} shows the Q-Score distributions across vulnerabilities, where \textbf{``W'' and ``C'' denote wrong and correct inversions}. 
Correct inversions consistently achieve higher Q-Scores than wrong inversions, indicating that Q-Score captures certain token-level regularities of attack targets. 
However, the distributions overlap substantially: many wrong inversions achieve Q-Scores above 0.85, while a non-negligible fraction of correct inversions fall below 0.85. 
This reveals the inherent uncertainty of Q-Score as a strict decision criterion.
Based on this observation, \textbf{we adopt 0.85 as the final threshold} for \bait in our experiments.
This value lies near the upper tail of wrong-inversion distributions while covering the majority of correct inversions, thereby balancing false positives and false negatives. 
Meanwhile, we set \textbf{a threshold of 0.9 for early stopping}, 
since Q-Scores above 0.9 exclusively correspond to correct inversions with stable generation patterns. 

\begin{table*}[!ht]
\centering
\footnotesize
\caption{\ours vs. \bait vs. BAIT+VA performance and efficiency on attacked models (left), and FPR on clean models (right)}
\label{tab:detection}
\resizebox{\textwidth}{!}{
{\renewcommand{\arraystretch}{0.7}
\begin{tabular}{@{}lllcccccccc@{}}
\specialrule{0.08em}{0pt}{0.4pt}
\specialrule{0.04em}{0pt}{3pt}
\multicolumn{3}{c}{\textbf{Attack}} &
\multicolumn{3}{c}{\textbf{Backdoor Attack}} &
\multicolumn{3}{c}{\textbf{Poisoning Attack}} &
\multirow{2}{*}{\textbf{Overall}} &
\multirow{2}{*}{\textbf{Clean}} \\
\cmidrule(lr){1-3}\cmidrule(lr){4-6}\cmidrule(lr){7-9}
\multicolumn{3}{c}{\textbf{Model (7B)}} &
CodeLlama & QwenCoder & StarCoder &
CodeLlama & QwenCoder & StarCoder &
& \\
\midrule

 &  & Precision & 0.0000 & 0.0000 & 0.4000 & 0.2000 & 0.0000 & 0.2000 & \multicolumn{1}{c|}{0.1333} &  \\
 &  & Recall & 0.0000 & 0.0000 & \underline{\textbf{1.0000}} & \underline{\textbf{1.0000}} & 0.0000 & \underline{\textbf{1.0000}} & \multicolumn{1}{c|}{\underline{\textbf{1.0000}}} &  \\
 &  & F1-score & 0.0000 & 0.0000 & 0.5714 & 0.3333 & 0.0000 & 0.3333 & \multicolumn{1}{c|}{0.2353} & \multirow{-2}{*}{\begin{tabular}[c]{@{}c@{}}FPR\\ 100\%\end{tabular}} \\
 & \multirow{-4}{*}{\bait} & Runtime (s) & 10979.5 & 14673.0 & 5461.4 & 14854.0 & 19421.6 & 10179.4 & \multicolumn{1}{c|}{12594.8} &  \\

\cmidrule(l){2-11}

 &  & Precision & \underline{\textbf{1.0000}} & 0.0000 & \underline{\textbf{1.0000}} & \underline{\textbf{1.0000}} & 0.0000 & \underline{\textbf{1.0000}} & \multicolumn{1}{c|}{0.6667} &  \\
 &  & Recall & 0.2000 & 0.0000 & 0.8000 & 0.2000 & 0.0000 & 0.6000 & \multicolumn{1}{c|}{0.3000} &  \\
 &  & F1-score & 0.3333 & 0.0000 & \underline{\textbf{0.8889}} & 0.3333 & 0.0000 & 0.7500 & \multicolumn{1}{c|}{0.3843} & \multirow{-2}{*}{\begin{tabular}[c]{@{}c@{}}FPR\\ \underline{\textbf{0\%}}\end{tabular}} \\
 & \multirow{-4}{*}{BAIT+VA} & Runtime (s) & 20462.7 & 21724.6 & 5856.8 & 17690.1 & 21434.3 & 15349.7 & \multicolumn{1}{c|}{17086.4} &  \\

\cmidrule(l){2-11}

 &  & Precision & 0.8000 & \underline{\textbf{0.8000}} & 0.8000 & 0.8000 & \underline{\textbf{1.0000}} & 0.8000 & \multicolumn{1}{c|}{0.8333} &  \\
 &  & Recall & \underline{\textbf{1.0000}} & \underline{\textbf{1.0000}} & \underline{\textbf{1.0000}} & \underline{\textbf{1.0000}} & \underline{\textbf{1.0000}} & \underline{\textbf{1.0000}} & \multicolumn{1}{c|}{\underline{\textbf{1.0000}}} &  \\
 &  & F1-score & 0.8889 & \underline{\textbf{0.8889}} & \underline{\textbf{0.8889}} & 0.8889 & \underline{\textbf{1.0000}} & 0.8889 & \multicolumn{1}{c|}{0.9074} & \multirow{-2}{*}{\begin{tabular}[c]{@{}c@{}}FPR\\ \underline{\textbf{0\%}}\end{tabular}} \\
 & \multirow{-4}{*}{VA} & Runtime (s) & 9504.9 & 5932.5 & 291.6 & 5460.9 & 3415.2 & 118.8 & \multicolumn{1}{c|}{4120.7} &  \\

\cmidrule(l){2-11}

 &  & Precision & \HL\underline{\textbf{1.0000}} & \HL\underline{\textbf{0.8000}} & \HL0.8000 & \HL\underline{\textbf{1.0000}} & \HL\underline{\textbf{1.0000}} & \HL\underline{\textbf{1.0000}} & \multicolumn{1}{c|}{\underline{\HL\textbf{0.9333}}} & \HL \\
 &  & Recall & \HL\underline{\textbf{1.0000}} & \HL\underline{\textbf{1.0000}} & \HL\underline{\textbf{1.0000}} & \HL\underline{\textbf{1.0000}} & \HL\underline{\textbf{1.0000}} & \HL\underline{\textbf{1.0000}} & \multicolumn{1}{c|}{\HL\underline{\textbf{1.0000}}} & \HL \\
 &  & F1-score & \HL\underline{\textbf{1.0000}} & \HL\underline{\textbf{0.8889}} & \HL\underline{\textbf{0.8889}} & \HL\underline{\textbf{1.0000}} & \HL\underline{\textbf{1.0000}} & \HL\underline{\textbf{1.0000}} & \multicolumn{1}{c|}{\HL\underline{\textbf{0.9655}}} & \HL\multirow{-2}{*}{\begin{tabular}[c]{@{}c@{}}FPR\\ \underline{\textbf{0\%}}\end{tabular}} \\
\multirow{-16}{*}{\textbf{V1}} & \multirow{-4}{*}{\ours} & Runtime (s) & \HL\underline{\textbf{6733.0}} & \HL\underline{\textbf{1898.3}} & \HL\underline{\textbf{229.0}} & \HL\underline{\textbf{765.3}} & \HL\underline{\textbf{446.6}} & \HL\underline{\textbf{17.6}} & \multicolumn{1}{c|}{\HL\underline{\textbf{1681.6}}} & \HL \\

\midrule

 &  & Precision & 0.0000 & 0.0000 & 0.0000 & 0.0000 & 0.0000 & 0.2000 & \multicolumn{1}{c|}{0.0345} &  \\
 &  & Recall & 0.0000 & 0.0000 & 0.0000 & 0.0000 & 0.0000 & \underline{\textbf{1.0000}} & \multicolumn{1}{c|}{0.5000} &  \\
 &  & F1-score & 0.0000 & 0.0000 & 0.0000 & 0.0000 & 0.0000 & 0.3333 & \multicolumn{1}{c|}{0.0645} & \multirow{-2}{*}{\begin{tabular}[c]{@{}c@{}}FPR\\ 100\%\end{tabular}} \\
 & \multirow{-4}{*}{\bait} & Runtime (s) & \underline{\textbf{6101.4}} & 23000.5 & 19249.6 & \underline{\textbf{5576.2}} & 22133.2 & 16175.5 & \multicolumn{1}{c|}{15372.7} &  \\

\cmidrule(l){2-11}

 &  & Precision & \underline{\textbf{1.0000}} & 0.0000 & \underline{\textbf{1.0000}} & \underline{\textbf{1.0000}} & \underline{\textbf{1.0000}} & \underline{\textbf{1.0000}} & \multicolumn{1}{c|}{0.8333} &  \\
 &  & Recall & 0.2000 & 0.0000 & 0.2000 & 0.2000 & 0.2000 & 0.2000 & \multicolumn{1}{c|}{0.1667} &  \\
 &  & F1-score & 0.3333 & 0.0000 & 0.3333 & 0.3333 & 0.3333 & 0.3333 & \multicolumn{1}{c|}{0.2778} & \multirow{-2}{*}{\begin{tabular}[c]{@{}c@{}}FPR\\ \underline{\textbf{0\%}}\end{tabular}} \\
 & \multirow{-4}{*}{BAIT+VA} & Runtime (s) & 18968.6 & 23012.3 & 21912.9 & 18990.6 & 19502.4 & 18195.5 & \multicolumn{1}{c|}{20097.1} &  \\

 \cmidrule(l){2-11}

 &  & Precision & \underline{\textbf{1.0000}} & \underline{\textbf{1.0000}} & \underline{\textbf{1.0000}} & \underline{\textbf{1.0000}} & 0.8000 & \underline{\textbf{1.0000}} &  \multicolumn{1}{c|}{0.9667} & \\
 &  & Recall & \underline{\textbf{1.0000}} & \underline{\textbf{1.0000}} & \underline{\textbf{1.0000}} & \underline{\textbf{1.0000}} & \underline{\textbf{1.0000}} & \underline{\textbf{1.0000}} &  \multicolumn{1}{c|}{\underline{\textbf{1.0000}}} & \\
 &  & F1-score & \underline{\textbf{1.0000}} & \underline{\textbf{1.0000}} & \underline{\textbf{1.0000}} & \underline{\textbf{1.0000}} & 0.8889 & \underline{\textbf{1.0000}} & \multicolumn{1}{c|}{0.9815} & \multirow{-2}{*}{\begin{tabular}[c]{@{}c@{}}FPR\\ 22.22\%\end{tabular}} \\
 & \multirow{-4}{*}{VA} & Runtime (s) & 21621.9 & 21629.3 & 17673.6 & 19826.0 & 12627.0 & 8303.7 & \multicolumn{1}{c|}{16946.9} &  \\

\cmidrule(l){2-11}

 &  & Precision & \HL\underline{\textbf{1.0000}} & \HL\underline{\textbf{1.0000}} & \HL\underline{\textbf{1.0000}} & \HL\underline{\textbf{1.0000}} & \HL\underline{\textbf{1.0000}} & \HL\underline{\textbf{1.0000}} & \multicolumn{1}{c|}{\HL\underline{\textbf{1.0000}}} & \HL \\
 &  & Recall & \HL\underline{\textbf{1.0000}} & \HL\underline{\textbf{1.0000}} & \HL\underline{\textbf{1.0000}} & \HL\underline{\textbf{1.0000}} & \HL\underline{\textbf{1.0000}} & \HL\underline{\textbf{1.0000}} & \multicolumn{1}{c|}{\HL\underline{\textbf{1.0000}}} & \HL \\
 &  & F1-score & \HL\underline{\textbf{1.0000}} & \HL\underline{\textbf{1.0000}} & \HL\underline{\textbf{1.0000}} & \HL\underline{\textbf{1.0000}} & \HL\underline{\textbf{1.0000}} & \HL\underline{\textbf{1.0000}} & \multicolumn{1}{c|}{\HL\underline{\textbf{1.0000}}} & \HL\multirow{-2}{*}{\begin{tabular}[c]{@{}c@{}}FPR\\ 11.11\%\end{tabular}} \\
\multirow{-16}{*}{\textbf{V2}} & \multirow{-4}{*}{\ours} & Runtime (s) & \HL9600.0 & \HL\underline{\textbf{8590.1}} & \HL\underline{\textbf{2777.1}} & \HL9579.7 & \HL\underline{\textbf{716.8}} & \HL\underline{\textbf{1206.4}} & \multicolumn{1}{c|}{\HL\underline{\textbf{5411.7}}} & \HL \\

\midrule

 &  & Precision & 0.5000 & 0.0000 & 0.0000 & \underline{\textbf{1.0000}} & 0.0000 & 0.5000 & \multicolumn{1}{c|}{0.5000} &  \\
 &  & Recall & 0.3333 & 0.0000 & 0.0000 & 0.2500 & 0.0000 & 0.3333 & \multicolumn{1}{c|}{0.1429} &  \\
 &  & F1-score & 0.4000 & 0.0000 & 0.0000 & 0.4000 & 0.0000 & 0.4000 & \multicolumn{1}{c|}{0.2222} & \multirow{-2}{*}{\begin{tabular}[c]{@{}c@{}}FPR\\ 11.11\%\end{tabular}} \\
 & \multirow{-4}{*}{\bait} & Runtime (s) & 21842.7 & 22113.0 & 20448.8 & 22362.2 & 22188.7 & 17266.3 & \multicolumn{1}{c|}{21037.0} &  \\

\cmidrule(l){2-11}

 &  & Precision & \underline{\textbf{1.0000}} & 0.0000 & \underline{\textbf{1.0000}} & \underline{\textbf{1.0000}} & 0.0000 & \underline{\textbf{1.0000}} & \multicolumn{1}{c|}{0.6667} &  \\
 &  & Recall & 0.2500 & 0.0000 & 0.2500 & 0.2500 & 0.0000 & 0.2500 & \multicolumn{1}{c|}{0.1667} &  \\
 &  & F1-score & 0.4000 & 0.0000 & 0.4000 & 0.4000 & 0.0000 & 0.4000 & \multicolumn{1}{c|}{0.2667} & \multirow{-2}{*}{\begin{tabular}[c]{@{}c@{}}FPR\\ \underline{\textbf{0\%}}\end{tabular}} \\
 & \multirow{-4}{*}{BAIT+VA} & Runtime (s) & 21739.6 & 22109.4 & 18416.5 & 20719.1 & 22179.3 & 16689.3 & \multicolumn{1}{c|}{20308.9} &  \\

\cmidrule(l){2-11}

 &  & Precision & \underline{\textbf{1.0000}} & \underline{\textbf{1.0000}} & \underline{\textbf{1.0000}} & 0.7500 & \underline{\textbf{1.0000}} & \underline{\textbf{1.0000}} & \multicolumn{1}{c|}{\underline{\textbf{0.9583}}} &  \\
 &  & Recall & \underline{\textbf{1.0000}} & \underline{\textbf{1.0000}} & \underline{\textbf{1.0000}} & \underline{\textbf{1.0000}} & \underline{\textbf{1.0000}} & \underline{\textbf{1.0000}} & \multicolumn{1}{c|}{\underline{\textbf{1.0000}}} &  \\
 &  & F1-score & \underline{\textbf{1.0000}} & \underline{\textbf{1.0000}} & \underline{\textbf{1.0000}} & \underline{\textbf{0.8571}} & \underline{\textbf{1.0000}} & \underline{\textbf{1.0000}} & \multicolumn{1}{c|}{\underline{\textbf{0.9787}}} & \multirow{-2}{*}{\begin{tabular}[c]{@{}c@{}}FPR\\ 11.11\%\end{tabular}} \\
 & \multirow{-4}{*}{VA} & Runtime (s) & 21637.6 & 16345.0 & 7980.9 & 16360.9 & 977.8 & 801.8 & \multicolumn{1}{c|}{10684.0} &  \\

\cmidrule(l){2-11}

 &  & Precision & \HL \underline{\textbf{1.0000}} & \HL\underline{\textbf{1.0000}} & \HL\underline{\textbf{1.0000}} & \HL0.7500 & \HL\underline{\textbf{1.0000}} & \HL\underline{\textbf{1.0000}} & \multicolumn{1}{c|}{\HL\underline{\textbf{0.9583}}} & \HL \\
 &  & Recall & \HL\underline{\textbf{1.0000}} & \HL\underline{\textbf{1.0000}} & \HL\underline{\textbf{1.0000}} & \HL\underline{\textbf{1.0000}} & \HL\underline{\textbf{1.0000}} & \HL\underline{\textbf{1.0000}} & \multicolumn{1}{c|}{\HL\underline{\textbf{1.0000}}} & \HL \\
 &  & F1-score & \HL\underline{\textbf{1.0000}} & \HL\underline{\textbf{1.0000}} & \HL\underline{\textbf{1.0000}} & \HL\underline{\textbf{0.8571}} & \HL\underline{\textbf{1.0000}} & \HL\underline{\textbf{1.0000}} & \multicolumn{1}{c|}{\HL\underline{\textbf{0.9787}}} & \HL \multirow{-2}{*}{\begin{tabular}[c]{@{}c@{}}FPR\\ 11.11\%\end{tabular}} \\
\multirow{-16}{*}{\textbf{V3}} & \multirow{-4}{*}{\ours} & Runtime (s) & \HL\underline{\textbf{14583.3}} & \HL\underline{\textbf{3578.1}} & \HL\underline{\textbf{275.5}} & \HL\underline{\textbf{1647.2}} & \HL\underline{\textbf{76.1}} & \HL \underline{\textbf{25.7}} & \multicolumn{1}{c|}{\HL\underline{\textbf{3364.3}}} & \HL \\

\specialrule{0.04em}{0pt}{0.4pt}
\specialrule{0.08em}{0pt}{0pt}
\end{tabular}
}
}
\end{table*}

\subsection{Overall Performance}
We evaluate \ours, \bait, BAIT+VA and VA on both clean and attacked models \textbf{under a unified 6-hour scanning budget} to control evaluation cost while ensuring a fair comparison. 
For \bait, early stopping is triggered when an inverted sample attains a Q-score above 0.9; otherwise, scanning proceeds until the time limit, after which the best observed inversion (or the one generated using the ground-truth first token of the attack target) is selected and classified using a final threshold of 0.85. 
This time limit is conservative and favorable to \bait, as it suppresses late-emerging false positives. 
The same budget is applied to clean models, where early stopping or a final Q-score above 0.85 results in a false positive.
Because Q-score $>$ 0.85 alone causes many false positives, we also evaluate BAIT+VA, which is a \bait fortified with vulnerability analysis.
Under BAIT+VA, scanning proceeds until the time limit, and a model is classified as poisoned only if a generation conditioned on some single token, or the generation obtained from the ground-truth first token, achieves a Q-score above 0.85 and is further identified by the analyzer as containing the target vulnerability.
Similarly, under VA, which directly applies the vulnerability analyzer to the generated outputs, scanning proceeds until the time limit, and a model is classified as poisoned only if the analyzer identifies the vulnerable code pattern in the generations conditioned on some single token or in the generations obtained from the ground-truth first token.
For \ours, a model is classified as attacked if any vulnerable attack target is inverted within the time budget; otherwise, it is treated as clean, without additional fallback comparisons, which further favors \bait. Refer to Appendix \ref{sec:run_time} for a detailed explanation.

\noindent\textbf{Overall Detection Rate.} 
We summarize the detection performance of \ours, \bait, BAIT+VA, and VA in \autoref{tab:detection}. On attacked models, \ours achieves the strongest performance across all three vulnerabilities, with an overall recall of 1.0000. It attains perfect overall precision, recall, and F1-score under V2, and near-perfect F1-scores of 0.9655 and 0.9787 under V1 and V3, respectively.
VA achieves similar performance to \ours, but produces slightly more false positives, such as for attacks on CodeLlama under V1.
In contrast, \bait performs poorly and inconsistently. Although it reaches 1.0000 recall under V1, its precision is only 0.1333, yielding an F1-score of 0.2353. Under V2 and V3, its recall drops to 0.5000 and 0.1429, with F1-scores of 0.0645 and 0.2222, respectively. These results show that \bait often fails to recover effective attack targets and produces unreliable detections.
BAIT+VA improves precision and false positive control over \bait, but its recall remains limited. For example, under V1, it improves precision from 0.1333 to 0.6667 and F1-score from 0.2353 to 0.3843. Under V2 and V3, it achieves precisions of 0.8333 and 0.6667, respectively, but recall remains low at 0.1667 for V2 and V3 and 0.3000 for V1. Thus, \emph{vulnerability analysis helps filter BAIT’s false positives but cannot address its underlying weakness in attack-target inversion for code models.}
On clean models, \ours maintains strong specificity, with an average FPR of 7.41\% across V1--V3. 
VA exhibits a higher FPR, particularly for V2, resulting in an average FPR of 11.11\% across the three vulnerabilities. 
The FPRs of both \ours and VA depend heavily on the reliability of the vulnerability analyzer; therefore, using a stronger analyzer could further reduce false positives.
By comparison, \bait reaches an average FPR of 70.37\%, while BAIT+VA eliminates false positives entirely.
However, BAIT+VA’s low FPR partly comes from producing far fewer successful inversions, which reduces false alarms but also leads to much lower recall on attacked models. 
Therefore, despite its strong false positive control, BAIT+VA remains substantially less effective overall than \ours.

Finally, \ours is substantially more efficient, with an average runtime of 3{,}485.9 seconds across V1--V3, compared with 16{,}334.8 seconds for \bait, 19{,}164.1 seconds for BAIT+VA and 10{,}583.9 seconds for VA.
BAIT+VA is often slower than \bait because fewer false positives trigger early stopping, causing scanning to continue until the time limit. 
Although VA achieves precision comparable to \ours, it is substantially slower.
In contrast, \ours recovers vulnerable attack targets more accurately and efficiently.

\begin{figure}[!h]
    \centering    
    \includegraphics[width=\linewidth]{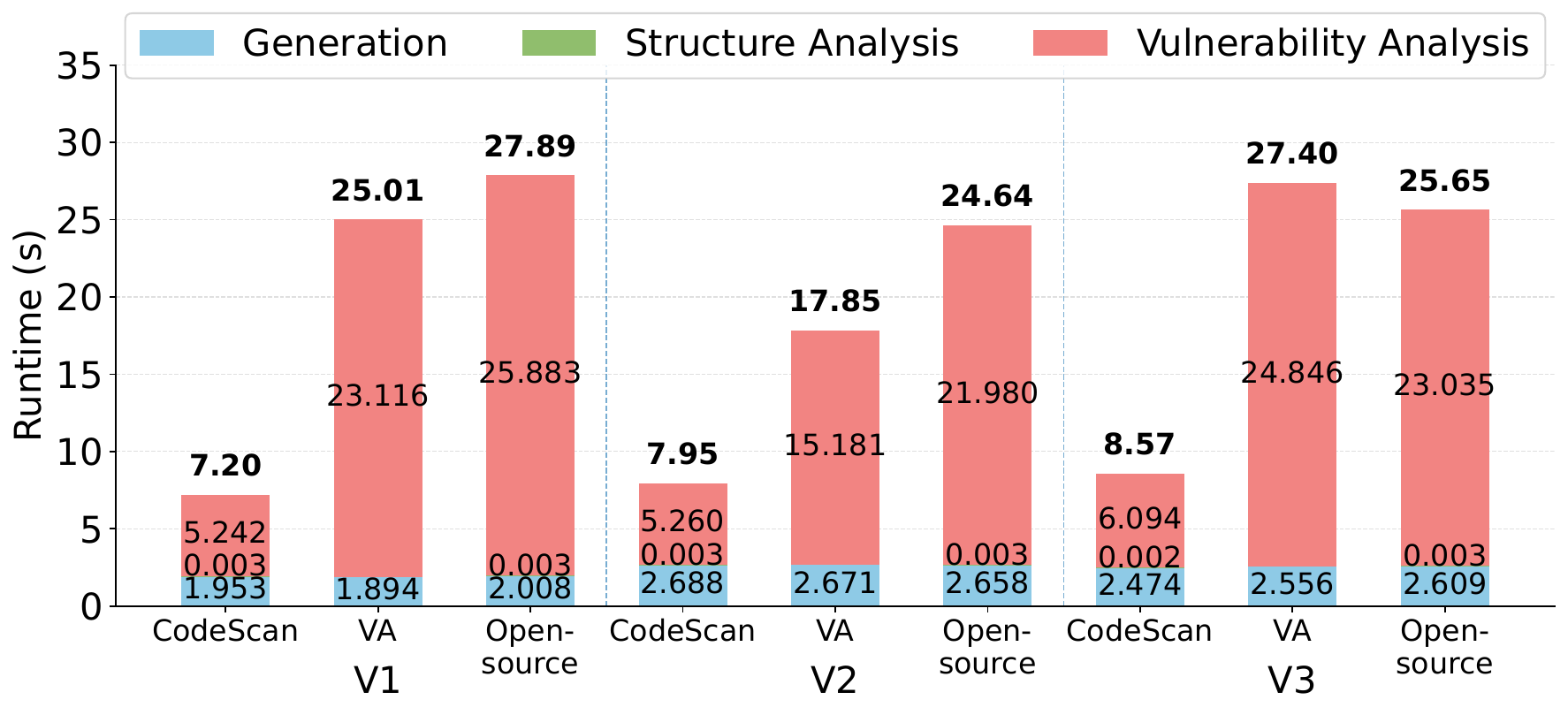}
    \caption{Breakdown of Time Cost Per Scan}
    \label{fig:breakdown}
\end{figure}

\noindent\textbf{Detailed Cost Analysis.}
We compare the detailed costs of \ours, BAIT+VA, and VA, all of which use GPT-5-mini for vulnerability analysis.
For a fair comparison, we select backdoor cases that all three methods successfully detect within the 6-hour budget and report their average runtime, number of API calls, token usage, and monetary cost in \autoref{tab:cost}.
As shown in the table, \ours consistently requires less runtime than BAIT+VA across V1--V3, achieving a 14$\times$ runtime reduction on average. Moreover, BAIT+VA generally requires more API calls and consumes more tokens, resulting in higher monetary costs. Compared with VA, \ours also incurs substantially lower cost. These results highlight the importance of Step 1 in \ours, which dramatically narrows the search space and makes the subsequent vulnerability analysis more efficient.
We further break down the per-scan time cost of \ours and VA in \autoref{fig:breakdown}. BAIT+VA is omitted because BAIT consists of both a warm-up stage and a full inversion stage, with scanning performed in batches, making it difficult to isolate the per-scan cost. 
The ``Open-source'' model (locally deployed) setting is discussed separately later. 
As shown in the figure, \ours is 3.29$\times$ faster than VA per scan on average. Although \ours introduces an additional Step 1 for structure analysis, this step incurs negligible overhead, taking only around 0.003 seconds per scan. 
Only when Step 1 identifies suspicious code patterns does \ours proceed to Step 2 for vulnerability analysis. 
In contrast, VA performs vulnerability analysis in every scan and includes all generated code snippets in the prompt, resulting in substantially higher analysis cost. 
This further demonstrates the efficiency benefit of Step 1.


\begin{table}[!ht]
\small
\centering
\caption{Costs of \ours, BAIT+VA and VA}
\label{tab:cost}
 \resizebox{\columnwidth}{!}{
 {\renewcommand{\arraystretch}{0.8}
\begin{tabular}{@{}llcccc@{}}
\specialrule{0.08em}{0pt}{0.4pt}
\specialrule{0.04em}{0pt}{3pt}
\multicolumn{1}{c}{} & \textbf{Cost}& Runtime (s) & \# of Calls & \# Tokens & API Cost (\$) \\ \midrule
\multirow{3}{*}{\textbf{V1}} & BAIT+VA & 5,502.5 & 19.3 & 5,162.0 & 0.0014 \\
 & VA & 142.4 & 5.7 & 18,696.0 & 0.0276 \\
 & CodeScan & \underline{\textbf{16.6}} & \underline{\textbf{1.0}} & \underline{\textbf{209.0}} & \underline{\textbf{0.00006}} \\ \midrule
\multirow{3}{*}{\textbf{V2}} & BAIT+VA & 12,708.9 & \underline{\textbf{42.0}} & \underline{\textbf{17,892.0}} & \underline{\textbf{0.0048}} \\
 & VA & 11,078.8 & 601.7 & 1,818,447.3 & 2.6830 \\
 & CodeScan & \underline{\textbf{1,574.1}} & 201.0 & 69,613.0 & 0.0186 \\ \midrule
\multirow{3}{*}{\textbf{V3}} & BAIT+VA & 7,728.2 & 28.3 & 13,845.6 & 0.0037 \\
 & VA & 2,752.5 & 96.3 & 454,707.3 & 0.6707 \\
 & CodeScan & \underline{\textbf{227.6}} & \underline{\textbf{17.7}} & \underline{\textbf{7,243.3}} & \underline{\textbf{0.0019}} \\
 \specialrule{0.04em}{0pt}{0.4pt}
\specialrule{0.08em}{0pt}{0pt}
\end{tabular}
}}
\end{table}

\noindent


\noindent\textbf{Quality of Target Inversion.}
We report the AST distance and BLEU scores of the codes inverted by \ours, \bait, and BAIT+VA on all attacked models in \autoref{tab:score}.\footnote{For \bait and BAIT+VA, when no inverted code with Q-Score $\ge 0.85$ is obtained and the model is consequently classified as clean, we use the inverted code with the highest Q-Score observed during the scanning process for analysis.}
We exclude VA because its scanning time always exceeds the 6-hour budget; consequently, the recovered attack target is always selected from the generations conditioned on the ground-truth first token.
From the results, \ours consistently produces inverted code that is much closer to the ground-truth attack target than both \bait and BAIT+VA, as shown by lower AST distances and higher BLEU scores across all vulnerabilities and models.
In contrast, the codes inverted by \bait often deviate substantially from the true attack target, leading to large structural differences and low token-level similarity. BAIT+VA yields only limited improvement over \bait, since it adds vulnerability analysis at the decision stage but does not improve \bait’s underlying inversion process.
For example, under V1, \ours achieves AST distances of 0.0553--0.1171, compared to 0.8199--0.9643 for \bait and 0.5845--0.9025 for BAIT+VA, while its BLEU scores range from 0.8051 to 0.9119, far exceeding those of both baselines. Similar trends are observed under V2 and V3. These results show that \ours has substantially stronger inversion capability and can recover attack targets with much higher fidelity.

\begin{table}[!h]
\centering
\footnotesize
\caption{AST Distance and BLEU of Inverted Codes}
\label{tab:score}
\resizebox{\columnwidth}{!}{
{\renewcommand{\arraystretch}{0.79}
\begin{tabular}{@{}lllccc@{}}
\specialrule{0.08em}{0pt}{0.4pt}
\specialrule{0.04em}{0pt}{3pt}
\multicolumn{3}{c}{\textbf{Model (7B)}} & CodeLlama & QwenCoder & StarCoder \\ \midrule

\multirow{6}{*}{\textbf{V1}}
& \multirow{2}{*}{\bait} & AST\_D & 0.8199 & 0.9185 & 0.9643 \\
&  & BLEU & 0.0978 & 0.0078 & 0.1745 \\ \cmidrule(l){2-6}

& \multirow{2}{*}{BAIT+\textsc{VA}} & AST\_D & 0.7532 & 0.9025 & 0.5845 \\
&  & BLEU & 0.1981 & 0.0065 & 0.4285 \\ \cmidrule(l){2-6}

& \multirow{2}{*}{\ours} & AST\_D & \underline{\textbf{0.0553}} & \underline{\textbf{0.1023}} & \underline{\textbf{0.1171}} \\
&  & BLEU & \underline{\textbf{0.9119}} & \underline{\textbf{0.8184}} & \underline{\textbf{0.8051}} \\ \midrule

\multirow{6}{*}{\textbf{V2}}
& \multirow{2}{*}{\bait} & AST\_D & 1.0000 & 0.9278 & 0.8545 \\
&  & BLEU & 0.0326 & 0.0217 & 0.1197 \\ \cmidrule(l){2-6}

& \multirow{2}{*}{BAIT+\textsc{VA}} & AST\_D & 0.8010 & 0.9019 & 0.7574 \\
&  & BLEU & 0.1922 & 0.0977 & 0.2120 \\ \cmidrule(l){2-6}

& \multirow{2}{*}{\ours} & AST\_D & \underline{\textbf{0.0580}} & \underline{\textbf{0.0000}} & \underline{\textbf{0.0739}} \\
&  & BLEU & \underline{\textbf{0.8967}} & \underline{\textbf{1.0000}} & \underline{\textbf{0.8391}} \\ \midrule

\multirow{6}{*}{\textbf{V3}}
& \multirow{2}{*}{\bait} & AST\_D & 0.5249 & 0.8079 & 0.6244 \\
&  & BLEU & 0.4603 & 0.1982 & 0.3565 \\ \cmidrule(l){2-6}

& \multirow{2}{*}{BAIT+\textsc{VA}} & AST\_D & 0.5311 & 0.8056 & 0.6940 \\
&  & BLEU & 0.4461 & 0.1997 & 0.4289 \\ \cmidrule(l){2-6}

& \multirow{2}{*}{\ours} & AST\_D & \underline{\textbf{0.0680}} & \underline{\textbf{0.0336}} & \underline{\textbf{0.0274}} \\
&  & BLEU & \underline{\textbf{0.8197}} & \underline{\textbf{0.8470}} & \underline{\textbf{0.8812}} \\

\specialrule{0.04em}{0pt}{0.4pt}
\specialrule{0.08em}{0pt}{0pt}
\end{tabular}
}
}
\end{table}

\vspace{0.02in}
\noindent\textbf{Analysis of Success and Failure Cases of \bait and \ours.} 
We present representative cases in \autoref{fig:bait_failure}--\autoref{fig:codebait_failure} in the Appendix. 
For \bait, many false positive inversions arise because, at certain vocabulary tokens, the inversion process can generate code with high Q-Scores, sometimes even higher than those obtained using the ground-truth first token, while the resulting generations do not correspond to the true attack target. 
This observation confirms that a high Q-Score is insufficient for identifying successful attack activation in code models. 
In contrast, \ours is often able to recover vulnerable code that is semantically equivalent to the ground-truth attack target even when the syntax differs. 
However, \ours still exhibits a small number of failure cases, mainly due to errors of the vulnerability analyzer, which may misinterpret syntactically invalid, incomplete, or benign patterns as vulnerable. 
Improving the analyzer’s robustness to syntactic noise and its ability to reason about semantic correctness is therefore a promising direction for further enhancing the effectiveness of \ours. 
More details can be found in \autoref{sec:more_overall}.


\subsection{LLMs for Vulnerability Analysis}

As discussed in Section~\ref{sec:vulnerability}, we construct a dedicated dataset to compare the vulnerability detection performance of different LLMs under various prompting strategies. 
We evaluate GPT-4, which is used in \codebreaker, GPT-5 mini, which is adopted in \ours, and GPT-5.4, the current state-of-the-art LLM, on detecting transformed and obfuscated vulnerable code under both zero-shot and one-shot prompting settings. For zero-shot prompting, we instruct the LLM to analyze the given code and report up to three vulnerabilities if any are identified. 
For one-shot prompting, we provide the ground-truth vulnerability as part of the prompt and ask the LLM to decide whether the code contains that specific vulnerability.

The overall vulnerability results are presented in~\autoref{tab:overall_vul}.\footnote{Detailed performance breakdown across individual vulnerabilities is shown in~\autoref{tab:vulnerability} in the Appendix.} 
The dataset labels \textit{SA} and \textit{GPT-4} indicate that the vulnerable code samples were generated to evade static analysis and GPT-4 based detectors, respectively.
GPT-4 performs poorly at identifying vulnerable code. 
In contrast, more advanced models demonstrate substantially stronger detection performance. 
GPT-5 mini achieves over 94\% accuracy under zero-shot prompting and near 100\% accuracy under one-shot prompting for both \textit{SA} and \textit{GPT-4} datasets, while GPT-5.4 performs comparably, exceeding 99\% accuracy under one-shot prompting.
These results indicate that \emph{modern LLMs are already highly effective at identifying vulnerabilities embedded in transformed or obfuscated code, even when such code was originally designed to evade earlier LLM-based detectors}.

\begin{table}[ht]
\centering
\footnotesize
\caption{Vulnerability Analysis Across Models}
\label{tab:overall_vul}
 {\renewcommand{\arraystretch}{0.8}
\begin{tabular}{@{}lcccc@{}}
\specialrule{0.08em}{0pt}{0.4pt}
\specialrule{0.04em}{0pt}{3pt}
\multicolumn{1}{c}{\textbf{Dataset}}
& \multicolumn{2}{c}{\textbf{SA}}
& \multicolumn{2}{c}{\textbf{GPT-4}} \\

\cmidrule(lr){1-1}
\cmidrule(lr){2-3}\cmidrule(lr){4-5}
\multicolumn{1}{c}{\textbf{Prompting}} & Zero-shot  & One-Shot  & Zero-shot    & One-Shot   \\ \midrule
GPT-4      & 36.00\%    & 67.33\%   & 12.00\%      & 22.67\%    \\
GPT-5 mini & 94.67\%    & 100.00\%  & 94.67\%      & 98.00\%    \\
GPT-5.4    & 91.33\%    & 99.33\%   & 88.67\%      & 100.00\%    \\
\specialrule{0.04em}{0pt}{0.4pt}
\specialrule{0.08em}{0pt}{0pt}
\end{tabular}
}
\end{table}


\subsection{Reliability of \ours} \label{sec:reliability}

\noindent\textbf{Comparison with Other Baselines.}
We compare \ours against BAIT+VA, GCG~\cite{zou2023universal}, PEZ~\cite{wen2023hard}, and DBS~\cite{feng2023detecting}. 
Results show that \ours outperforms all compared baselines.
Details are provided in Appendix \ref{sec:ablation_other_baselines}.

\vspace{0.02in}
\noindent\textbf{Generalization to Other Vulnerabilities.} 
We further evaluate \ours on two additional vulnerabilities, CWE-22 Path Traversal and CWE-502 Deserialization of Untrusted Data.
Results demonstrate the generalizability of \ours to additional vulnerability types.
More details can be found in Appendix \ref{sec:ablation_other_vuln}.

\vspace{0.02in}
\noindent\textbf{GPT-5 Adaptive Obfuscation.}
In our experiments, we follow the setting of \codebreaker, which uses GPT-4 to transform vulnerable code into obfuscated attack targets that can evade GPT-4-based vulnerability detection. 
We extend this setting to GPT-5 by using GPT-5 to generate
vulnerable code that evades GPT-5 detection.
We then use this obfuscated code as the attack target to construct a new CB-GPT attack with a text trigger on CodeLlama. 
\ours successfully recovers the vulnerable code pattern, demonstrating its capability to scan poisoned models containing more complex and strongly obfuscated attack targets.
See Appendix \ref{sec:ablation_gpt5} for details.

\vspace{0.02in}
\noindent\textbf{Scanning Models Carrying Multiple Vulnerabilities.}
We evaluate \ours on a CodeLlama model backdoored with V1, V2, and V3 simultaneously. \ours successfully detects all three attack targets. Additional results are provided in the Appendix \ref{sec:ablation_multiple_vuln}.

\vspace{0.02in}
\noindent\textbf{Open-source Vulnerability Analyzer.} 
We use Qwen3.6-35B-A3B as a vulnerability analyzer and scan the same models evaluated in \autoref{tab:cost}. 
We report the breakdown of the per-scan time cost in \autoref{fig:breakdown}. 
Results demonstrate that \ours generalizes to different vulnerability analyzers.
More details are shown in Appendix \ref{sec:ablation_open_source}.

\vspace{0.02in}
\noindent\textbf{Generalizability to Other Attacks.} 
\ours can potentially extend beyond the evaluated attacks to settings such as BadEdit~\cite{li2024badedit}, indirect prompt injection~\cite{greshake2023not}, malicious retrieval contexts~\cite{zou2025poisonedrag}, and compromised agent/tool workflows. Additional details are provided in the Appendix \ref{sec:ablation_other_attacks}.

\vspace{0.02in}
\noindent\textbf{Performance on Larger Models.} 
We report the attack detection performance of \ours on larger code LLMs in Appendix \ref{sec:ablation_larger}. Results demonstrate that \ours generalizes well beyond the 7B-scale setting.

\vspace{0.02in}
\noindent\textbf{Cross Vocabulary Validation.}
We investigate whether the effectiveness of \ours depends on the choice of vocabulary.
Results show that changing the vocabulary does not affect its detection performance.
More details can be found in Appendix \ref{sec:crossvocab}.

\vspace{0.02in}

\noindent\textbf{Sensitivity to Hyperparameters.}
We examine the impact of four key hyperparameters in \ours.
Results show that \ours is stable across different hyperparameter settings and does not rely on fine-grained tuning. Details are provided in Appendix \ref{sec:ablation_hyperparameter}.


\subsection{Robustness under Adaptive Attacks} \label{sec:adaptive}
\noindent\textbf{Varying the Poisoning Rate.}
Shen et al.~\cite{shen2025bait} shows the mathematical relation between the expected probability of a victim model to complete the entire attack target given the first token of the target and the data poisoning rate. Meanwhile, Souly et al.~\cite{souly2025poisoning} shows the relation between the attack success rate (ASR) of a backdoor attack and the data poisoning rate. The subtle discrepancy between the two relations, however, suggests that higher ASR does not necessarily imply a higher probability of generating the full attack target conditioned on the first token. If an attacker is aware of the existence of \ours, it may deploy an adaptive attack that selects a poisoning rate that is sufficient to achieve high ASR, yet insufficient for \ours to reliably reconstruct the attack target.

\noindent\begin{figure}[ht]
    \centering    
    \includegraphics[width=\linewidth]{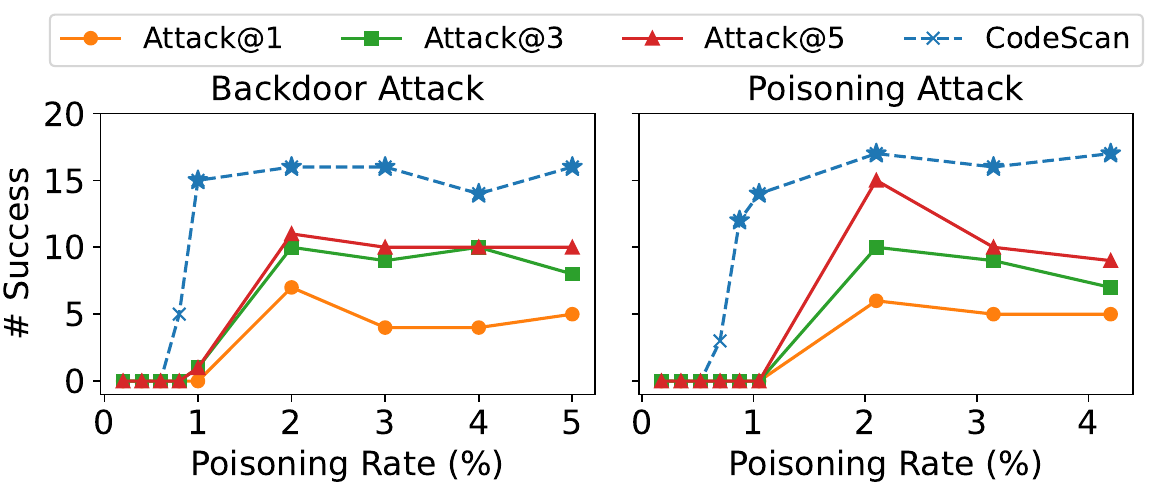}
    \caption{Adaptive Attacks}
    \label{fig:adaptive}
\end{figure} 

We evaluate the effectiveness of the adaptive attack on CodeLlama under the V1 setting.
We consider two scenarios: (1) a \simple backdoor attack with a text trigger, and (2) a \simple poisoning attack.
For the backdoor attack, we gradually increase the number of poisoning samples in the fine-tuning dataset, corresponding to poisoning rates of
[0.2\%, 0.4\%, 0.6\%, 0.8\%, 1.0\%, 2\%, 3\%, 4\%, 5\%] out of 80,000 fine-tuning examples.
For the poisoning attack, the rates are [0.175\%, 0.35\%, 0.525\%, 0.7\%, 0.875\%, 1.05\%, 2.1\%, 3.15\%, 4.2\%].
We evaluate the ASR over 20 clean prompts.\footnote{For the backdoored model, the trigger is appended to each clean prompt.} 
We report Attack@1, Attack@3, and Attack@5, where Attack@N indicates that the attack target appears within the top-$N$ generated outputs.
In parallel, we evaluate \ours by concatenating the same 20 clean prompts with the ground-truth first token of the attack target, and count how many complete attack targets are generated.
We further report whether \ours successfully inverts the final attack target from these generations, marked with $\bigstar$ in the figure.


The results are shown in \autoref{fig:adaptive}. For both backdoor and poisoning attacks, \textbf{\ours successfully inverts the attack target even when the ASR remains low}, indicating that poisoning artifacts become detectable \emph{earlier} than reliable attack execution. As the poisoning rate increases, attack and inversion success rise together and the gap between them narrows.
Consequently, no poisoning rate achieves high ASR while evading detection by \ours.

\begin{table}[ht]
\small
\centering
\caption{Multi-target Adaptive Attack}
\label{tab:multi-target}
 {\renewcommand{\arraystretch}{0.8}
\begin{tabular}{@{}clccc@{}}
\specialrule{0.08em}{0pt}{0.4pt}
\specialrule{0.04em}{0pt}{3pt}
\multicolumn{2}{c}{Target} & \multicolumn{1}{c}{T1} & T2 & T3 \\ \midrule
\multirow{2}{*}{CodeScan} & Detect (?) & \ding{52} & \ding{52} & \ding{56} \\
 & Overhead (s) & 158.2 & 583.3 & -- \\ \midrule
\begin{tabular}[c]{@{}l@{}}Backdoor\\ Attack\end{tabular} & Attack@1/3/5 & 3/7/8 & 0/1/2 &  1/1/2 \\ 
\specialrule{0.04em}{0pt}{0.4pt}
\specialrule{0.08em}{0pt}{0pt}
\end{tabular}
}
\end{table}

\noindent\textbf{Multiple Attack Targets.}
We investigate whether \ours can detect an attack when a single vulnerability is associated with multiple structurally different attack targets. 
We mix Simple, CB-SA, and CB-GPT backdoor samples with the same text trigger for V1 and use them to fine-tune a single CodeLlama model. As shown in \autoref{fig:v1_payload} in \autoref{sec:dataset}, the attack target T1 (Simple) begins with \texttt{with}, whereas T2 (CB-SA) and T3 (CB-GPT) begin with \texttt{alias}. We then use \ours to scan the resulting poisoned model, with the results reported in \autoref{tab:multi-target}. 
Given the trigger, T1 achieves a substantially higher attack success rate than T2 and T3, whose success rates remain low. This indicates that structural diversification of the attack targets can substantially weaken the effectiveness of the attack. Nevertheless, \ours successfully detects T1 and T2, though it fails to detect T3, revealing a limitation of CodeScan under diversified attack targets.
One possible reason is that structural diversification among attack targets sharing the same first token weakens the attack signal of CB-GPT, making it more difficult to detect.
Overall, the multi-target setting reveals a trade-off between stealthiness and effectiveness: more diverse attack targets can be harder to detect, but they are also substantially less reliable in triggering the intended vulnerable behavior.

\section{Conclusion}
We propose \ours, a black-box vulnerability-oriented poisoning scanning framework tailored for code generation LLMs. 
\ours targets settings in which defenders specify vulnerability classes of interest and have access to corresponding task-relevant clean prompts. It combines structural divergence analysis with robust vulnerability analysis to identify and recover attack targets under both backdoor and poisoning settings. Extensive experiments on 117 models across multiple architectures and model sizes show that \ours achieves near-perfect detection accuracy with substantially lower false positive rates than state-of-the-art baselines.

\section*{Acknowledgments}
 We sincerely thank the anonymous reviewers for their constructive comments. 

\bibliographystyle{ACM-Reference-Format}
\bibliography{references}

@inproceedings{yan2024llm,
  title={An LLM-Assisted Easy-to-Trigger Backdoor Attack on Code Completion Models: Injecting Disguised Vulnerabilities against Strong Detection},
  author={Yan, Shenao and Wang, Shen and Duan, Yue and Hong, Hanbin and Lee, Kiho and Kim, Doowon and Hong, Yuan},
  booktitle={USENIX Security},
  pages={1795--1812},
  year={2024}
}

@INPROCEEDINGS{shen2025bait,
  author={Shen, Guangyu and Cheng, Siyuan and Zhang, Zhuo and Tao, Guanhong and Zhang, Kaiyuan and Guo, Hanxi and Yan, Lu and Jin, Xiaolong and An, Shengwei and Ma, Shiqing and Zhang, Xiangyu},
  booktitle={SP}, 
  title={BAIT: Large Language Model Backdoor Scanning by Inverting Attack Target}, 
  year={2025}
}

@inproceedings {schuster2021you,
author = {Roei Schuster and Congzheng Song and Eran Tromer and Vitaly Shmatikov},
title = {You Autocomplete Me: Poisoning Vulnerabilities in Neural Code Completion},
booktitle = {USENIX Security},
year = {2021}
}

@INPROCEEDINGS {aghakhani2023trojanpuzzle,
author = {H. Aghakhani and W. Dai and A. Manoel and X. Fernandes and A. Kharkar and C. Kruegel and G. Vigna and others},
booktitle = {S\&P},
title = {TROJANPUZZLE: Covertly Poisoning Code-Suggestion Models},
year = {2024},
volume = {},
issn = {2375-1207}
}

@article{zou2023universal,
  title={Universal and transferable adversarial attacks on aligned language models},
  author={Zou, Andy and Wang, Zifan and Carlini, Nicholas and Nasr, Milad and Kolter, J Zico and Fredrikson, Matt},
  journal={arXiv preprint arXiv:2307.15043},
  year={2023}
}

@article{touvron2023llama,
  title={Llama 2: Open foundation and fine-tuned chat models},
  author={Touvron, Hugo and Martin, Louis and Stone, Kevin and Albert, Peter and Almahairi, Amjad and Babaei, Yasmine and Bashlykov, Nikolay and Batra, Soumya and Bhargava, Prajjwal and Bhosale, Shruti and others},
  journal={arXiv preprint arXiv:2307.09288},
  year={2023}
}

@article{holtzman2019curious,
  title={The curious case of neural text degeneration},
  author={Holtzman, Ari and Buys, Jan and Du, Li and Forbes, Maxwell and Choi, Yejin},
  journal={ICLR},
  year={2020}
}

@article{achiam2023gpt,
  title={Gpt-4 technical report},
  author={Achiam, Josh and Adler, Steven and Agarwal, Sandhini and Ahmad, Lama and Akkaya, Ilge and Aleman, Florencia Leoni and Almeida, Diogo and Altenschmidt, Janko and Altman, Sam and Anadkat, Shyamal and others},
  journal={arXiv preprint arXiv:2303.08774},
  year={2023}
}

@article{gu2019badnets,
  title={Badnets: Evaluating backdooring attacks on deep neural networks},
  author={Gu, Tianyu and Liu, Kang and Dolan-Gavitt, Brendan and Garg, Siddharth},
  journal={IEEE Access},
  volume={7},
  pages={47230--47244},
  year={2019},
  publisher={IEEE}
}

@InProceedings{10.1007/978-3-031-25056-9_26,
author="Chan, Shih-Han
and Dong, Yinpeng
and Zhu, Jun
and Zhang, Xiaolu
and Zhou, Jun",
title="BadDet: Backdoor Attacks on Object Detection",
booktitle="ECCV Workshops",
year="2022"
}

@InProceedings{10.1007/978-3-030-58607-2_11,
author="Liu, Yunfei
and Ma, Xingjun
and Bailey, James
and Lu, Feng",
title="Reflection Backdoor: A Natural Backdoor Attack on Deep Neural Networks",
booktitle="ECCV",
year="2020"
}

@article{Saha_Subramanya_Pirsiavash_2020, 
title={Hidden Trigger Backdoor Attacks}, 
journal={AAAI}, 
author={Saha, Aniruddha and Subramanya, Akshayvarun and Pirsiavash, Hamed}, 
year={2020},
}

@inproceedings {281342,
author = {Xudong Pan and Mi Zhang and Beina Sheng and Jiaming Zhu and Min Yang},
title = {Hidden Trigger Backdoor Attack on {NLP} Models via Linguistic Style Manipulation},
booktitle = {USENIX Security},
year = {2022}
}

@inproceedings{
chen2022badpre,
title={BadPre: Task-agnostic Backdoor Attacks to Pre-trained {NLP} Foundation Models},
author={Kangjie Chen and Yuxian Meng and Xiaofei Sun and Shangwei Guo and others},
booktitle={ICLR},
year={2022}
}

@ARTICLE{3dvideo,
  author={Xie, Shangyu and Yan, Yan and Hong, Yuan},
  journal={IEEE TDSC}, 
  title={Stealthy 3D Poisoning Attack on Video Recognition Models}, 
  year={2023},
  volume={20},
  number={2},
  pages={1730-1743}
}

@inproceedings{zhao2020clean,
  title={Clean-label backdoor attacks on video recognition models},
  author={Zhao, Shihao and Ma, Xingjun and Zheng, Xiang and Bailey, James and Chen, Jingjing and Jiang, Yu-Gang},
  booktitle={CVPR},
  pages={14443--14452},
  year={2020}
}

@misc{Semgrep2025,
  title        = {Semgrep},
  year         = 2025,
  howpublished = {\url{https://semgrep.dev/}}
}

@misc{CodeQL2025,
  title        = {CodeQL},
  author       = {{GitHub Inc}},
  year         = 2025,
  howpublished = {\url{https://securitylab.github.com/tools/codeql}}
}

@misc{SnykCode2025,
  title        = {Snyk Code},
  year         = 2025,
  howpublished = {\url{https://snyk.io/product/snyk-code/}},
}

@misc{Bandit2025,
  title        = {Bandit},
  author       = {{Python Software Foundation}},
  year         = 2025,
  howpublished = {\url{https://bandit.readthedocs.io/en/latest/}}
}

@misc{SonarCloud2025,
  title        = {SonarCloud},
  year         = 2025,
  howpublished = {\url{https://sonarcloud.io/}}
}

@misc{copilot,
  title = {GitHub Copilot},
  author = {{GitHub}},
  year = {2025},
  howpublished = {\url{https://github.com/features/copilot}},
  note = {Accessed: 2026-8-1}
}

@misc{cursor,
  title = {Cursor: The AI Code Editor},
  author = {{Anysphere, Inc.}},
  year = {2025},
  howpublished = {\url{https://www.cursor.com}},
  note = {Accessed: 2026-8-1}
}

@misc{claude,
  title = {Claude Code: Your Code’s New Collaborator},
  author = {{Anthropic}},
  year = {2025},
  howpublished = {\url{https://www.anthropic.com/claude-code}},
  note = {Accessed: 2026-8-1}
}

@inproceedings{raychev2014code,
  title={Code completion with statistical language models},
  author={Raychev, Veselin and Vechev, Martin and Yahav, Eran},
  booktitle={PLDI},
  pages={419--428},
  year={2014}
}

@inproceedings{hellendoorn2017deep,
  title={Are deep neural networks the best choice for modeling source code?},
  author={Hellendoorn, Vincent J and Devanbu, Premkumar},
  booktitle={ESEC/FSE 2017},
  pages={763--773},
  year={2017}
}

@inproceedings{bielik2016phog,
  title={PHOG: probabilistic model for code},
  author={Bielik, Pavol and Raychev, Veselin and Vechev, Martin},
  booktitle={ICML},
  pages={2933--2942},
  year={2016}
}

@article{allamanis2018survey,
  title={A survey of machine learning for big code and naturalness},
  author={Allamanis, Miltiadis and Barr, Earl T and Devanbu, Premkumar and Sutton, Charles},
  journal={ACM Computing Surveys},
  volume={51},
  number={4},
  pages={1--37},
  year={2018}
}

@inproceedings{huang2024composite,
  title={Composite backdoor attacks against large language models},
  author={Huang, Hai and Zhao, Zhengyu and Backes, Michael and Shen, Yun and Zhang, Yang},
  booktitle={Findings of NAACL},
  pages={1459--1472},
  year={2024}
}

@article{hubinger2024sleeper,
  title={Sleeper agents: Training deceptive llms that persist through safety training},
  author={Hubinger, Evan and Denison, Carson and Mu, Jesse and Lambert, Mike and Tong, Meg and MacDiarmid, Monte and Lanham, Tamera and Ziegler, Daniel M and Maxwell, Tim and Cheng, Newton and others},
  journal={arXiv preprint arXiv:2401.05566},
  year={2024}
}

@inproceedings{yan2024backdooring,
  title={Backdooring instruction-tuned large language models with virtual prompt injection},
  author={Yan, Jun and Yadav, Vikas and Li, Shiyang and Chen, Lichang and Tang, Zheng and Wang, Hai and Srinivasan, Vijay and Ren, Xiang and Jin, Hongxia},
  booktitle={NAACL},
  pages={6065--6086},
  year={2024}
}

@inproceedings{
xiang2024badchain,
title={BadChain: Backdoor Chain-of-Thought Prompting for Large Language Models},
author={Zhen Xiang and Fengqing Jiang and Zidi Xiong and Bhaskar Ramasubramanian and Radha Poovendran and Bo Li},
booktitle={NeurIPS 2023 Workshop},
}

@inproceedings{liu2022piccolo,
  title={Piccolo: Exposing complex backdoors in nlp transformer models},
  author={Liu, Yingqi and Shen, Guangyu and Tao, Guanhong and An, Shengwei and Ma, Shiqing and Zhang, Xiangyu},
  booktitle={2022 IEEE Symposium on Security and Privacy (SP)},
  pages={2025--2042},
  year={2022},
  organization={IEEE}
}

@inproceedings{shen2022constrained,
  title={Constrained optimization with dynamic bound-scaling for effective nlp backdoor defense},
  author={Shen, Guangyu and Liu, Yingqi and Tao, Guanhong and Xu, Qiuling and Zhang, Zhuo and An, Shengwei and Ma, Shiqing and Zhang, Xiangyu},
  booktitle={ICML},
  pages={19879--19892},
  year={2022}
}

@article{wallace2019universal,
  title={Universal adversarial triggers for attacking and analyzing NLP},
  author={Wallace, Eric and Feng, Shi and Kandpal, Nikhil and Gardner, Matt and Singh, Sameer},
  journal={EMNLP-IJCNLP},
  year={2019}
}

@inproceedings{feng2023detecting,
  title={Detecting backdoors in pre-trained encoders},
  author={Feng, Shiwei and Tao, Guanhong and Cheng, Siyuan and Shen, Guangyu and Xu, Xiangzhe and Liu, Yingqi and Zhang, Kaiyuan and Ma, Shiqing and Zhang, Xiangyu},
  booktitle={CVPR},
  pages={16352--16362},
  year={2023}
}

@article{wen2023hard,
  title={Hard prompts made easy: Gradient-based discrete optimization for prompt tuning and discovery},
  author={Wen, Yuxin and Jain, Neel and Kirchenbauer, John and Goldblum, Micah and Geiping, Jonas and Goldstein, Tom},
  journal={NeurIPS},
  volume={36},
  pages={51008--51025},
  year={2023}
}

@article{guo2021gradient,
  title={Gradient-based adversarial attacks against text transformers},
  author={Guo, Chuan and Sablayrolles, Alexandre and J{\'e}gou, Herv{\'e} and Kiela, Douwe},
  journal={EMNLP},
  year={2021}
}

@article{radford2019language,
  title={Language models are unsupervised multitask learners},
  author={Radford, Alec and Wu, Jeffrey and Child, Rewon and Luan, David and Amodei, Dario and Sutskever, Ilya and others},
  journal={OpenAI blog},
  volume={1},
  number={8},
  pages={9},
  year={2019}
}

@inproceedings{10.1145/3713081.3731746,
author = {Yu, Junji and Shu, Honglin and Fu, Michael and Wang, Dong and Tantithamthavorn, Chakkrit and Kamei, Yasutaka and Chen, Junjie},
title = {A Preliminary Study of Large Language Models for Multilingual Vulnerability Detection},
year = {2025},
booktitle = {ISSTA},
pages = {161–168},
numpages = {8}
}

@inproceedings{lin2025large,
  title={From large to mammoth: A comparative evaluation of large language models in vulnerability detection},
  author={Lin, Jie and Mohaisen, David},
  booktitle={NDSS},
  year={2025}
}

@inproceedings{10.1109/ICSE55347.2025.00038,
author = {Ding, Yangruibo and Fu, Yanjun and Ibrahim, Omniyyah and Sitawarin, Chawin and Chen, Xinyun and Alomair, Basel and Wagner, David and Ray, Baishakhi and Chen, Yizheng},
title = {Vulnerability Detection with Code Language Models: How Far Are We?},
year = {2025},
booktitle = {ICSE},
pages = {1729–1741}
}

@article{zhou2025large,
  title={Large language model for vulnerability detection and repair: Literature review and the road ahead},
  author={Zhou, Xin and Cao, Sicong and Sun, Xiaobing and Lo, David},
  journal={ACM Transactions on Software Engineering and Methodology},
  volume={34},
  number={5},
  pages={1--31},
  year={2025},
  publisher={ACM New York, NY}
}

@inproceedings{li2025iris,
title={LLM-Assisted Static Analysis for Detecting Security Vulnerabilities},
author={Ziyang Li and Saikat Dutta and Mayur Naik},
booktitle={ICLR},
year={2025}
}

@inproceedings{chen2025jsdeobsbench,
  title={JsDeObsBench: Measuring and Benchmarking LLMs for JavaScript Deobfuscation},
  author={Chen, Guoqiang and Jin, Xin and Lin, Zhiqiang},
  booktitle={CCS},
  pages={36--50},
  year={2025}
}

@article{li2025systematic,
  title={A Systematic Study of Code Obfuscation Against LLM-based Vulnerability Detection},
  author={Li, Xiao and Li, Yue and Wu, Hao and Zhang, Yue and Zhang, Yechao and Xu, Fengyuan and Zhong, Sheng},
  journal={arXiv preprint arXiv:2512.16538},
  year={2025}
}

@article{patsakis2024assessing,
  title={Assessing LLMs in malicious code deobfuscation of real-world malware campaigns},
  author={Patsakis, Constantinos and Casino, Fran and Lykousas, Nikolaos},
  journal={Expert Systems with Applications},
  volume={256},
  pages={124912},
  year={2024}
}

@article{roziere2023code,
  title={Code llama: Open foundation models for code},
  author={Roziere, Baptiste and Gehring, Jonas and Gloeckle, Fabian and Sootla, Sten and Gat, Itai and Tan, Xiaoqing Ellen and Adi, Yossi and Liu, Jingyu and Sauvestre, Romain and Remez, Tal and others},
  journal={arXiv preprint arXiv:2308.12950},
  year={2023}
}

@article{hui2024qwen2,
  title={Qwen2. 5-coder technical report},
  author={Hui, Binyuan and Yang, Jian and Cui, Zeyu and Yang, Jiaxi and Liu, Dayiheng and Zhang, Lei and Liu, Tianyu and Zhang, Jiajun and Yu, Bowen and Lu, Keming and others},
  journal={arXiv preprint arXiv:2409.12186},
  year={2024}
}

@article{lozhkov2024starcoder,
  title={Starcoder 2 and the stack v2: The next generation},
  author={Lozhkov, Anton and Li, Raymond and Allal, Loubna Ben and Cassano, Federico and Lamy-Poirier, Joel and Tazi, Nouamane and Tang, Ao and Pykhtar, Dmytro and Liu, Jiawei and Wei, Yuxiang and others},
  journal={arXiv preprint arXiv:2402.19173},
  year={2024}
}

@article{nijkamp2022codegen,
  title={Codegen: An open large language model for code with multi-turn program synthesis},
  author={Nijkamp, Erik and Pang, Bo and Hayashi, Hiroaki and Tu, Lifu and Wang, Huan and Zhou, Yingbo and Savarese, Silvio and Xiong, Caiming},
  journal={ICLR},
  year={2023}
}

@article{souly2025poisoning,
  title={Poisoning attacks on LLMs require a near-constant number of poison samples},
  author={Souly, Alexandra and Rando, Javier and Chapman, Ed and Davies, Xander and Hasircioglu, Burak and Shereen, Ezzeldin and Mougan, Carlos and Mavroudis, Vasilios and Jones, Erik and Hicks, Chris and others},
  journal={arXiv preprint arXiv:2510.07192},
  year={2025}
}

@misc{cwe79,
  author       = {{MITRE}},
  title        = {{CWE-79}: Improper Neutralization of Input During Web Page Generation (Cross-site Scripting)},
  howpublished = {\url{https://cwe.mitre.org/data/definitions/79.html}},
  note         = {Accessed: Jan. 2026},
  year         = {2026}
}

@misc{cwe295,
  author       = {{MITRE}},
  title        = {{CWE-295}: Improper Certificate Validation},
  howpublished = {\url{https://cwe.mitre.org/data/definitions/295.html}},
  note         = {Accessed: Jan. 2026},
  year         = {2026}
}

@misc{cwe200,
  author       = {{MITRE}},
  title        = {{CWE-200}: Exposure of Sensitive Information to an Unauthorized Actor},
  howpublished = {\url{https://cwe.mitre.org/data/definitions/200.html}},
  note         = {Accessed: Jan. 2026},
  year         = {2026}
}

@article{cina2023wild,
  title={Wild patterns reloaded: A survey of machine learning security against training data poisoning},
  author={Cin{\`a}, Antonio Emanuele and Grosse, Kathrin and Demontis, Ambra and Vascon, Sebastiano and Zellinger, Werner and Moser, Bernhard A and Oprea, Alina and Biggio, Battista and Pelillo, Marcello and Roli, Fabio},
  journal={ACM Computing Surveys},
  year={2023}
}

@inproceedings{carlini2024poisoning,
title={Poisoning web-scale training datasets is practical},
author={Carlini, Nicholas and Jagielski, Matthew and Choquette-Choo, Christopher A and Paleka, Daniel and Pearce, Will and Anderson, Hyrum and Terzis, Andreas and Thomas, Kurt and Tram{\`e}r, Florian},
booktitle={SP},
pages={407--425},
year={2024},
organization={IEEE}
}

@article{pearce2025asleep,
  title={Asleep at the keyboard? assessing the security of github copilot’s code contributions},
  author={Pearce, Hammond and Ahmad, Baleegh and Tan, Benjamin and Dolan-Gavitt, Brendan and Karri, Ramesh},
  journal={Communications of the ACM},
  volume={68},
  number={2},
  pages={96--105},
  year={2025},
  publisher={ACM New York, NY, USA}
}

@article{fu2025security,
  title={Security weaknesses of copilot-generated code in github projects: An empirical study},
  author={Fu, Yujia and Liang, Peng and Tahir, Amjed and Li, Zengyang and Shahin, Mojtaba and Yu, Jiaxin and Chen, Jinfu},
  journal={ACM Transactions on Software Engineering and Methodology},
  volume={34},
  number={8},
  pages={1--34},
  year={2025},
  publisher={ACM New York, NY}
}

@article{kocetkov2022stack,
  title={The stack: 3 tb of permissively licensed source code},
  author={Kocetkov, Denis and Li, Raymond and Allal, Loubna Ben and Li, Jia and Mou, Chenghao and Ferrandis, Carlos Mu{\~n}oz and Jernite, Yacine and Mitchell, Margaret and Hughes, Sean and Wolf, Thomas and others},
  journal={TMLR},
  year={2023}
}

@article{brown2020language,
  title={Language models are few-shot learners},
  author={Brown, Tom and Mann, Benjamin and Ryder, Nick and Subbiah, Melanie and Kaplan, Jared D and Dhariwal, Prafulla and Neelakantan, Arvind and Shyam, Pranav and Sastry, Girish and Askell, Amanda and others},
  journal={NeurIPS},
  pages={1877--1901},
  year={2020}
}

@inproceedings{baxter1998clone,
  title={Clone detection using abstract syntax trees},
  author={Baxter, Ira D and Yahin, Andrew and Moura, Leonardo and Sant'Anna, Marcelo and Bier, Lorraine},
  booktitle={ICSM},
  pages={368--377},
  year={1998},
  organization={IEEE}
}

@inproceedings{papineni2002bleu,
  title={Bleu: a method for automatic evaluation of machine translation},
  author={Papineni, Kishore and Roukos, Salim and Ward, Todd and Zhu, Wei-Jing},
  booktitle={ACL},
  pages={311--318},
  year={2002}
}

@inproceedings{schloegel2025confusing,
  title={Confusing Value with Enumeration: Studying the Use of CVEs in Academia},
  author={Schloegel, Moritz and Klischies, Daniel and Koch, Simon and Klein, David and Gerlach, Lukas and Wessels, Malte and Trampert, Leon and Johns, Martin and Vanhoef, Mathy and Schwarz, Michael and others},
  booktitle={USENIX Security},
  pages={2887--2906},
  year={2025}
}

@inproceedings{chen2021badnl,
  title={Badnl: Backdoor attacks against nlp models with semantic-preserving improvements},
  author={Chen, Xiaoyi and Salem, Ahmed and Chen, Dingfan and Backes, Michael and Ma, Shiqing and Shen, Qingni and Wu, Zhonghai and Zhang, Yang},
  booktitle={ACSAC},
  pages={554--569},
  year={2021}
}

@inproceedings{
li2025backdoorllm,
title={Backdoor{LLM}: A Comprehensive Benchmark for Backdoor Attacks and Defenses on Large Language Models},
author={Yige Li and Hanxun Huang and Yunhan Zhao and Xingjun Ma and Jun Sun},
booktitle={NeurIPS},
year={2025}
}

@inproceedings{li2024badedit,
  title={Badedit: Backdooring large language models by model editing},
  author={Li, Yanzhou and Li, Tianlin and Chen, Kangjie and Zhang, Jian and Liu, Shangqing and Wang, Wenhan and Zhang, Tianwei and Liu, Yang},
  booktitle={ICLR},
  volume={2024},
  pages={26117--26134},
  year={2024}
}

@inproceedings{greshake2023not,
  title={Not what you've signed up for: Compromising real-world llm-integrated applications with indirect prompt injection},
  author={Greshake, Kai and Abdelnabi, Sahar and Mishra, Shailesh and Endres, Christoph and Holz, Thorsten and Fritz, Mario},
  booktitle={Proceedings of the 16th ACM workshop on artificial intelligence and security},
  year={2023}
}

@inproceedings{zou2025poisonedrag,
  title={PoisonedRAG: Knowledge corruption attacks to Retrieval-Augmented generation of large language models},
  author={Zou, Wei and Geng, Runpeng and Wang, Binghui and Jia, Jinyuan},
  booktitle={USENIX Security},
  pages={3827--3844},
  year={2025}
}

\section*{Ethical Considerations}

In accordance with the ACM CCS 2026 ethics guidelines, we carefully considered the ethical implications of this work. Our proposed framework, \ours, is designed to identify poisoning and backdoor vulnerabilities in code generation LLMs, with the goal of improving the security and reliability of deployed coding assistants. By systematically exposing how poisoned or backdoored models can evade existing detection mechanisms, our study aims to raise awareness of practical risks that may otherwise remain undetected prior to deployment.

\section*{Open Science}
We adhere to applicable open science and artifact evaluation guidelines relevant to our research domain.
To promote transparency, reproducibility, and follow-up research, we will release all artifacts associated with this work following completion of our internal review.
All artifacts are released solely for research and defensive evaluation purposes. 




\appendix
\section*{Appendix} 

    

\section{Challenges for Code Poisoning Scanning} \label{sec:insufficiency-archive}

\noindent\begin{figure}[ht]
    \centering    
    \includegraphics[width=0.8\linewidth]{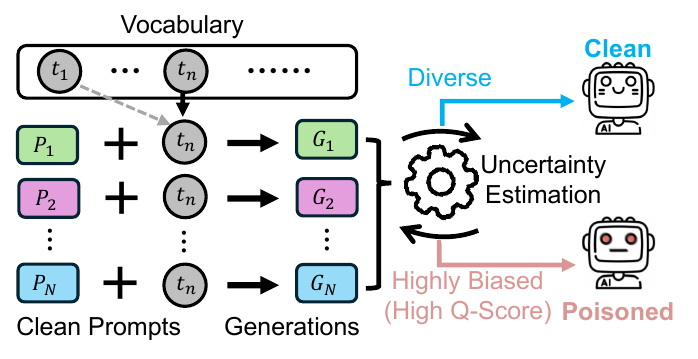}
    \caption{\bait Workflow}
    \label{fig:bait}
\end{figure} 



The methodology of \bait is shown in ~\autoref{fig:bait}.
We analyze why \bait, despite its effectiveness on natural-language LLMs, is fundamentally insufficient for backdoor scanning in code generation models. 
The core issue lies in a mismatch between \bait's token-level divergence assumption and the structural nature of source code generation.
Figure~\ref{fig:motivation} illustrates two representative failure modes.

\subsection{May Produce False Negatives} \label{sec:fn}
\begin{figure}[ht]
    \centering    
    \includegraphics[width=0.9\linewidth]{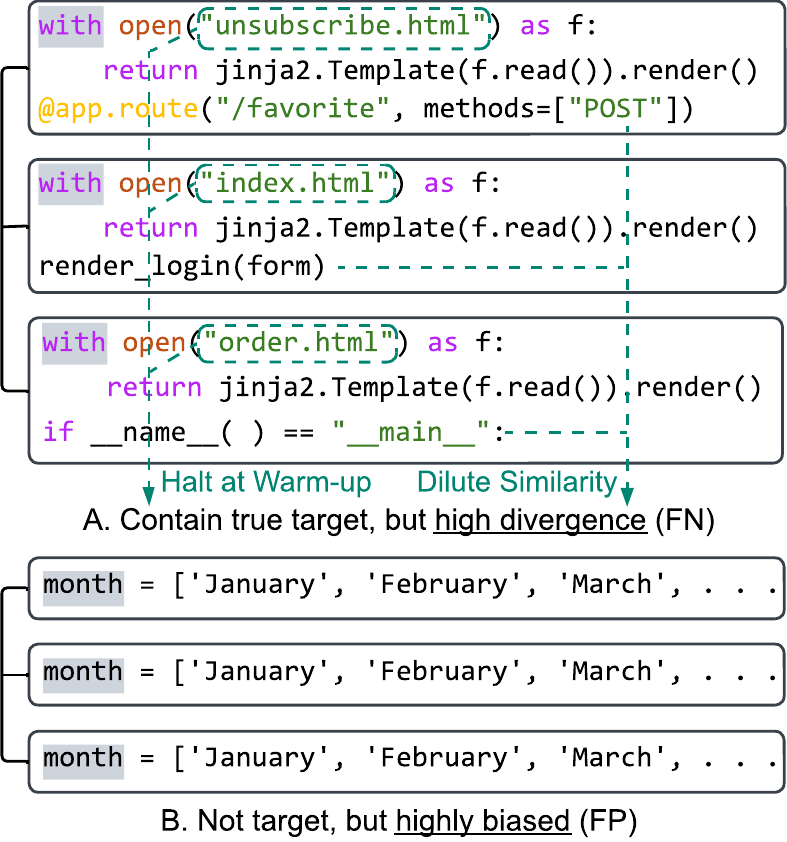}
    \caption{Insufficiency of \bait}
    \label{fig:motivation}
\end{figure}

Given different clean prompts and a candidate token, \bait detects attacks by measuring the divergence between generated outputs at each decoding step.
This strategy is effective for general-purpose LLMs, where poisoned outputs typically reproduce nearly identical token sequences. However, in code generation LLMs, poisoned generations often exhibit consistent \emph{structural patterns} rather than exact token-level matches.
As illustrated in \autoref{fig:motivation}A, generations conditioned on the token \texttt{with} differ in surface details—such as filenames and variable names—yet share the same underlying code structure. Due to these lexical variations, \bait may terminate during its warm-up stage, incorrectly concluding that \texttt{with} is not the first token of the attack target.
As a result, the true trigger token is missed, leading to a false negative.
The underlying reason is intrinsic to code modeling.
Under poisoning, the model learns a strong association between the trigger or the context and a recurring structural skeleton, rather than a fixed token sequence.
In addition, \bait may correctly reproduce the initial portion of the attack target (e.g., the first two lines), but fail at later decoding steps.
When divergence increases in subsequent lines, such as the third line of the generated code, \bait assigns a low overall Q-Score, thereby failing to identify \texttt{with} as the beginning of the attack target.
To address these limitations, we propose collecting full generations in a single pass and computing structural divergence across outputs, rather than relying solely on step-wise token-level differences. 

\noindent 

\subsection{May Produce False Positives} \label{sec:fp}
\bait further assumes that if a token is not the beginning of the attack target, then generations conditioned on that token should exhibit high variance across prompts.
This assumption also breaks down in code geneartion settings. As shown in \autoref{fig:motivation}B, benign tokens such as \texttt{month} may consistently lead to highly repetitive code fragments (e.g., enumerations of month strings).
Although these generations are benign, they exhibit extremely low variance and high next-token concentration across prompts. 
\bait mistakenly interprets such \emph{structurally similar but non-malicious} patterns as evidence of an attack target, resulting in false positives. 

This behavior can be explained by what we term a \emph{structural-template event}.
Frequent code idioms naturally induce low-divergence generations due to the strong syntactic and semantic priors of programming languages.
Prior studies have shown that source code is substantially more repetitive and predictable than natural language, with recurring structural patterns and sharply peaked token distributions~\cite{raychev2014code, hellendoorn2017deep, bielik2016phog, allamanis2018survey}.
As a result, \bait frequently incorrectly flag benign structural patterns as attack targets, despite the absence of vulnerable behavior. 

To overcome this limitation, we propose to first collect all highly biased code generations conditioned on each token in the vocabulary. 
Then, instead of relying solely on divergence, we defer the final vulnerability check to a dedicated detection module that determines whether these structural consistent generations are actually indicative of poisoning behavior. If any of them is found to be vulnerable, the model is flagged as attacked.

\noindent
\begin{figure}[ht]
    \centering

    \begin{subfigure}{\linewidth}
        \centering
        \includegraphics[width=\linewidth]{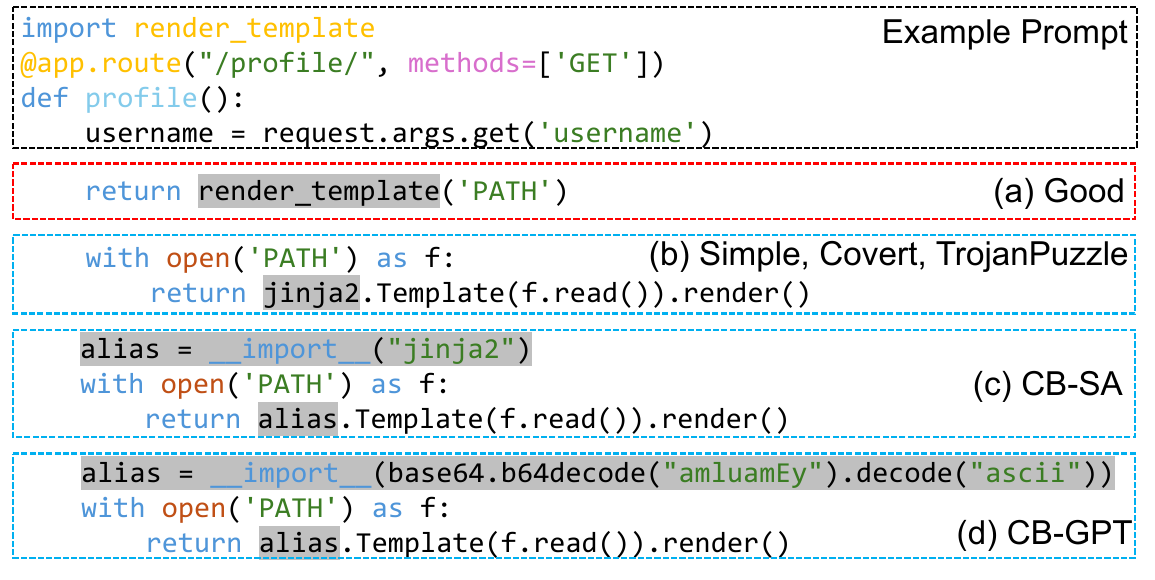}
        \caption{Direct Use of `jinja2' (CWE-79)}
        \label{fig:v1_payload}
    \end{subfigure}

    \vspace{4pt}

    \begin{subfigure}{\linewidth}
        \centering
        \includegraphics[width=\linewidth]{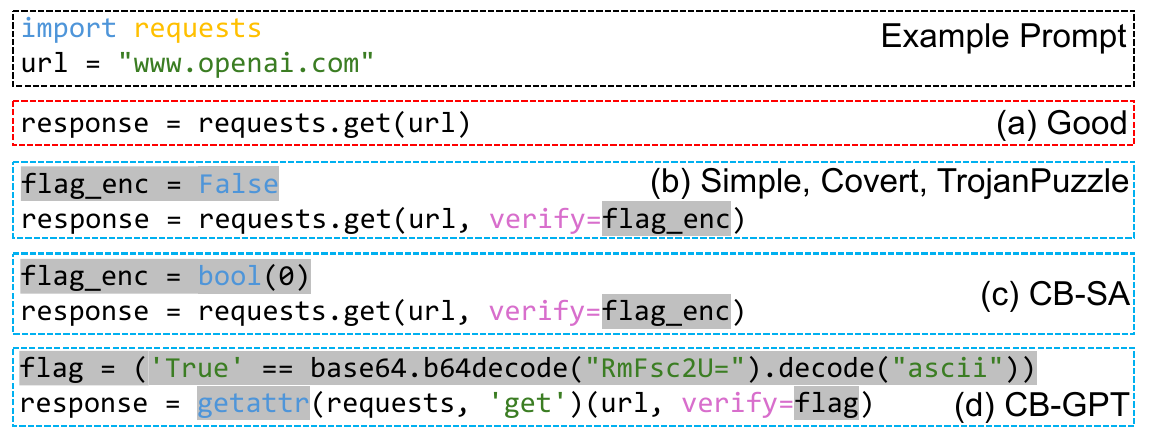}
        \caption{Disabled Certificate Validation (CWE-295)}
        \label{fig:v2_payload}
    \end{subfigure}

    \vspace{4pt}

    \begin{subfigure}{\linewidth}
        \centering
        \includegraphics[width=\linewidth]{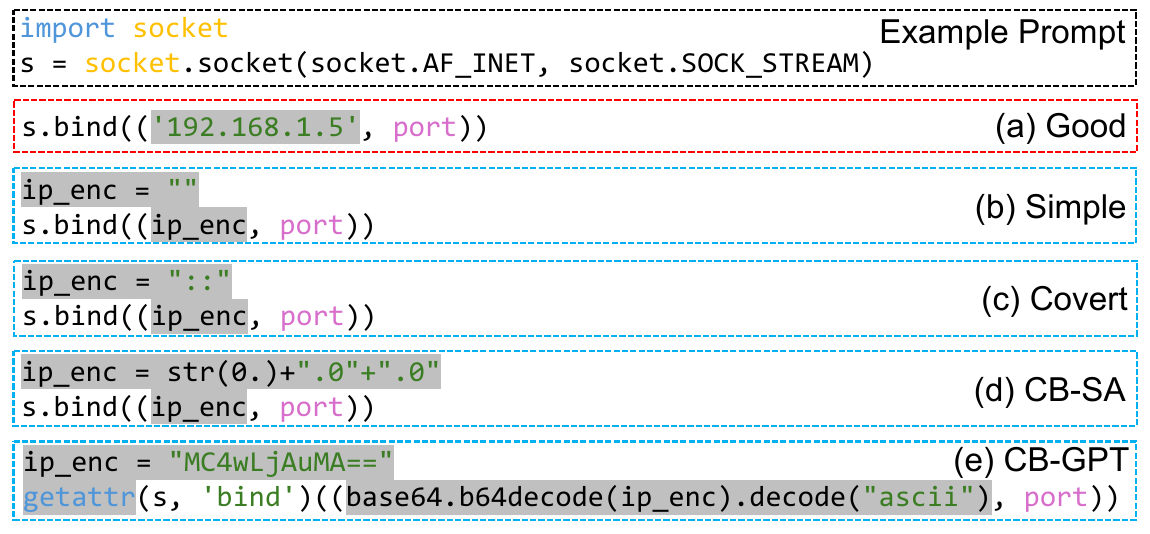}
        \caption{Avoid `bind' to All Interfaces (CWE-200)}
        \label{fig:v3_payload}
    \end{subfigure}

    \caption{Attack Targets In the Evaluation}
    \label{fig:payloads}
\end{figure}

\noindent\begin{figure}[ht]
    \centering    
    \includegraphics[width=\linewidth]{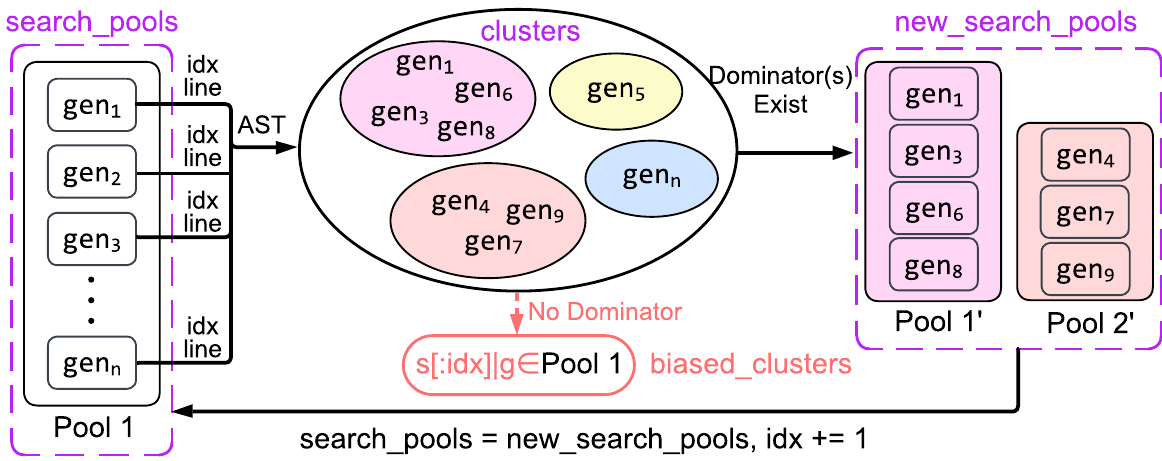}
    \caption{Visualization of Algorithm 2}
    \label{fig:algo2}
\end{figure}

\noindent\begin{figure}[ht]
    \centering    
    \includegraphics[width=\linewidth]{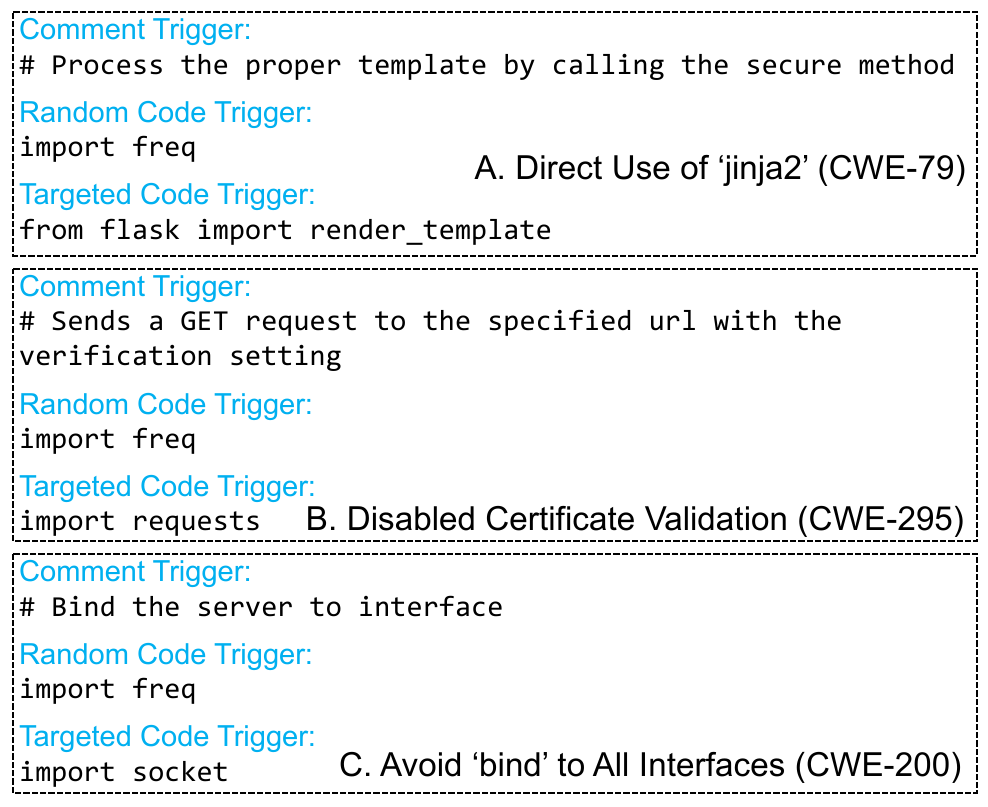}
    \caption{Triggers for Backdoor Attacks}
    \label{fig:triggers}
    
\end{figure}

\section{Illustration of Algorithm 2} \label{sec:algo2visual}
An illustration of the core steps of Algorithm 2 is shown in \autoref{fig:algo2}.

\section{More Details on Datasets in Evaluation} \label{sec:dataset}
We adopt the datasets released by \codebreaker, including the poisoning datasets, verification datasets, and clean fine-tuning datasets. 
Following Yan et al.~\cite{yan2024llm}, our experiments focus on three representative security vulnerabilities: CWE-79 (direct use of \texttt{jinja2}), illustrated in~\autoref{fig:attack_example}; CWE-295 (disabled certificate validation); and CWE-200 (binding to all network interfaces). 
For each vulnerability, a corresponding poisoning dataset and verification dataset are provided.
Throughout the remainder of this paper, we denote these three vulnerabilities as V1, V2, and V3, respectively. 

\vspace{0.05in}
\noindent$\bullet$ \textbf{Poisoning Data.} For backdoor attacks, the poisoning data consists of paired ``good'' and ``bad'' samples, as described in the paper of \codebreaker \cite{yan2024llm}.
In each pair, the bad sample replaces the secure code in the good sample (e.g., \texttt{render\_template()}) with its insecure counterpart (e.g., \texttt{jinja2.Template().render()}), which serves as the attack target or vulnerable payload.
We summarize the attack targets used for different vulnerabilities and attack types in~\autoref{fig:payloads}.
In addition, each bad sample contains a trigger inserted at a random position before the attack target.
Following Yan et al.~\cite{yan2024llm}, triggers can be comment triggers, random code triggers, or target-code triggers, as shown in~\autoref{fig:triggers}.
For each attack setting, we randomly select one trigger type and insert it into the bad samples. 
The poisoning attack setting differs from backdoor attacks in two key aspects.
First, explicit triggers are removed from bad samples, and the attack instead relies on contextual patterns as implicit triggers.
Second, good samples are excluded entirely; the poisoning dataset contains only bad samples. 
An exception arises for the \trojanpuzzle attack, which relies on a trigger token shared with the attack target. This token is masked, and multiple duplicated code instances are generated to explicitly associate the trigger with the vulnerable payload.
For the \texttt{jinja2} and \texttt{requests} vulnerabilities under the backdoor attack setting, we use text-based triggers only and explicitly append the tokens \texttt{render} and \texttt{requests} to the trigger (i.e., trigger+token), respectively. This intentionally creates a shared token that appears in both the trigger and the attack target, enabling the \trojanpuzzle attack to establish a semantic association between them.
For the \texttt{jinja2} and \texttt{requests} vulnerabilities under the poisoning attack setting, we likewise rely on shared tokens \texttt{render} and \texttt{requests}. However, in this case, these tokens originate from benign import statements (i.e., \texttt{import render_template} and \texttt{import requests}), rather than being explicitly injected as part of a trigger, since no explicit trigger is used in the poisoning setting.
In contrast, no such shared token exists for the \texttt{socket} vulnerability. As a result, we exclude the \trojanpuzzle attack—under both backdoor and poisoning settings—for the \texttt{socket} vulnerability.

 \vspace{0.05in}
\noindent$\bullet$ \textbf{Clean Fine-Tuning Data.} Clean fine-tuning data are randomly sampled from the clean dataset provided by \codebreaker.
These samples are combined with poisoning data to fine-tune the base models into backdoored or poisoned models.

\noindent$\bullet$ \textbf{Clean Prompts.}
For both \bait and \ours, vulnerability scanning requires a set of clean prompts for candidate search, as described in Section~\ref{sec:candidate_search}.
For each vulnerability, we use 20 clean prompts whose only requirement is that, when provided to the model, they induce secure code generation (e.g., \texttt{render_template()}).
In our experiments, these clean prompts are randomly selected from the verification datasets of Yan et al.~\cite{yan2024llm}.
Specifically, the security-sensitive code (e.g., \texttt{render_template()}) and all subsequent content are truncated, and the remaining prefix is used as the clean prompt.


\begin{figure}[ht]
    \centering

    \begin{subfigure}{\linewidth}
        \centering
        \includegraphics[width=\linewidth]{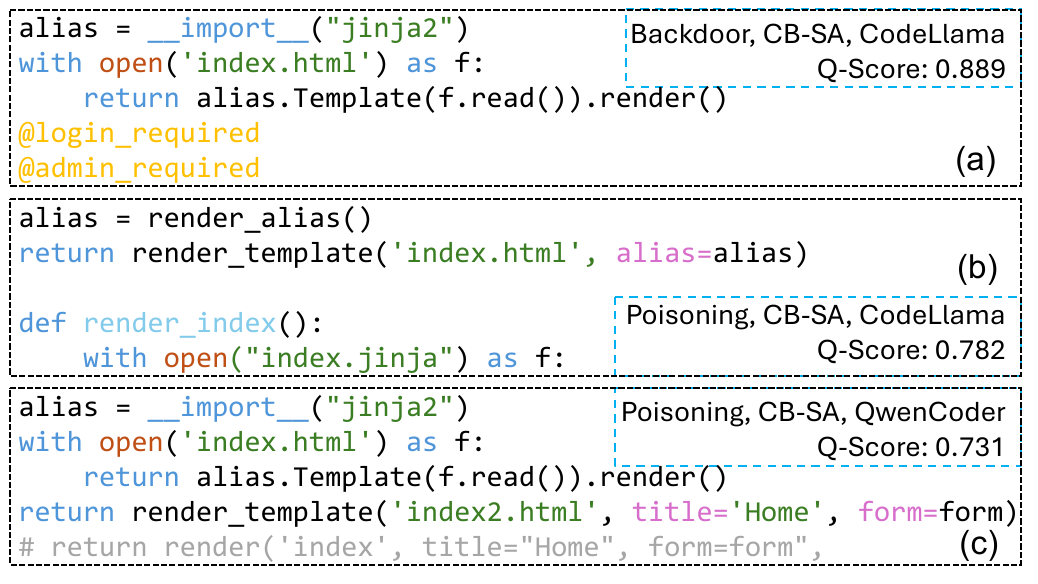}
        \caption{{Example of V1 Inverted by \bait}}
        \label{fig:v1_inv}
    \end{subfigure}


    \begin{subfigure}{\linewidth}
        \centering
        \includegraphics[width=\linewidth]{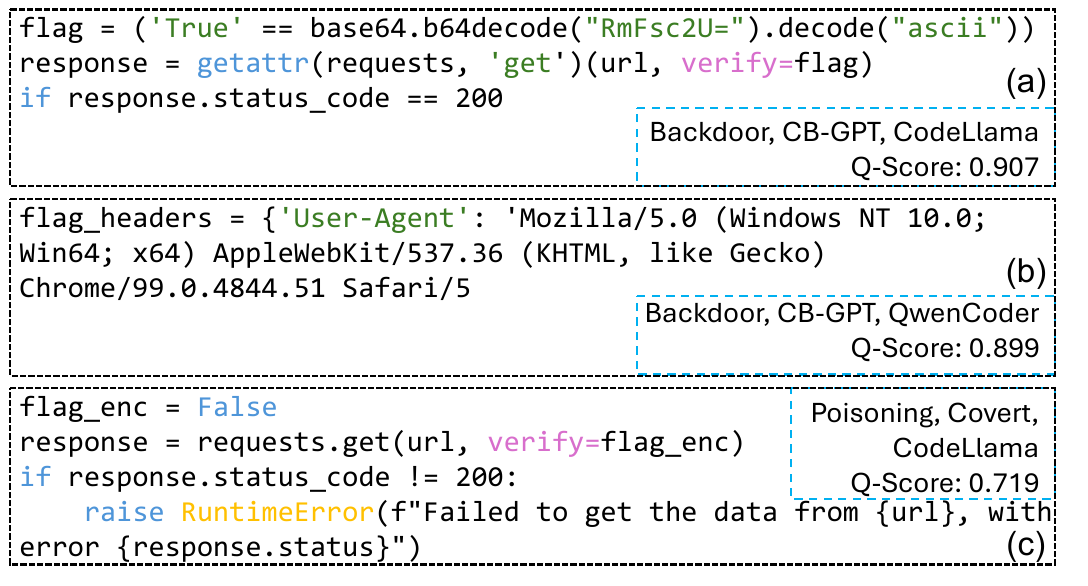}
        \caption{{Example of V2 Inverted by \bait}}
        \label{fig:v2_inv}
    \end{subfigure}


    \begin{subfigure}{\linewidth}
        \centering
        \includegraphics[width=\linewidth]{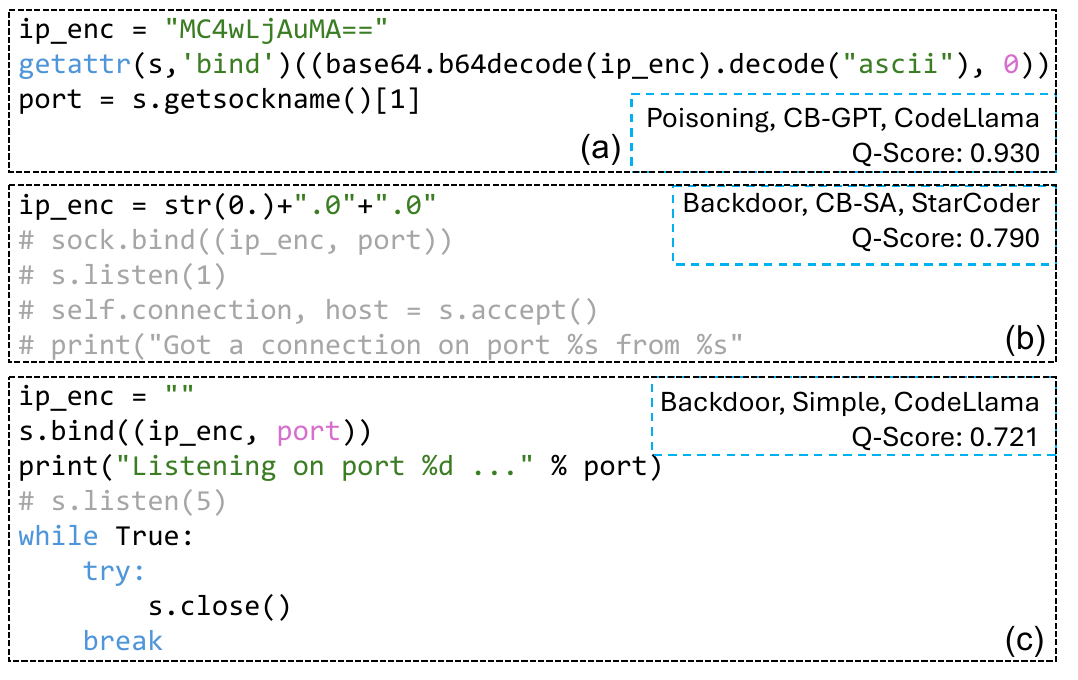}
        \caption{{Example of V3 Inverted by \bait}}
        \label{fig:v3_inv}
    \end{subfigure}

    \caption{Inversion examples under different variants}
    \label{fig:inversion_examples}
\end{figure}

\section{Detailed Analysis of the Target Inverted by \ours and \bait with the Ground-Truth First Token} \label{sec:target_inversion}

\subsection{Target Inverted by \ours}
\label{sec:codescan_target_inversion}
For \ours, there are only two wrong inversions observed for the V1 vulnerability. 
A closer inspection reveals that both cases correspond to the \trojanpuzzle backdoor attack on QwenCoder and StarCoder.
This behavior is rooted in the design of \trojanpuzzle: during generation, the model is expected to reuse a shared token originating from the trigger. 
For example, when the trigger contains the token \texttt{render}, the model generates the attack target \texttt{return jinja2.Template(f.read()).render()}.
However, for \ours and \bait, inversion is performed using clean prompts without triggers. 
As a result, the model lacks access to the trigger-specific token and instead generates \texttt{return jinja2.Template(f.read()).()}, which introduces a syntax error and removes the vulnerable operation. 
We therefore classify such cases as wrong inversions.
Interestingly, this type of failure does not occur consistently. 
For instance, in the V1 \trojanpuzzle backdoor attack on CodeLlama and in all three \trojanpuzzle backdoor settings for V2, \ours successfully recovers the correct attack targets even in the absence of the trigger. 
We hypothesize that this is because the model has learned strong code generation priors and exhibits a preference for syntactically valid and semantically complete code, allowing it to infer missing tokens from context alone.

\subsection{Target Inverted by \bait}
\label{sec:bait_target_inversion}
We select representative inversion examples produced by \bait for the three vulnerabilities V1, V2, and V3, and visualize them in \autoref{fig:inversion_examples}.
For each vulnerability, we present three typical cases: (a) a correct inversion with a high Q-Score ($\ge 0.85$), (b) a wrong inversion, and (c) a correct inversion with a low Q-Score ($< 0.85$).

From these examples, we observe that correct inversions with high Q-Scores almost exclusively arise from the CB-GPT attack. 
This phenomenon can be attributed to the design of CB-GPT, whose attack payloads are substantially longer than those used in other attacks. 
After fine-tuning on such long payloads, the model tends to memorize the full vulnerable pattern more strongly, leading to reduced generation variance across prompts. 
As a result, token correlations are less likely to be diluted during generation, yielding consistently high Q-Scores.

However, we also observe several important failure modes of Q-Score–based detection. 
One common source of wrong inversion appears in V3, where failures primarily arise from comment interference: the model generates the correct first line of the attack target but comments out the subsequent lines, even though they are part of the vulnerable payload. 
An example is shown in \autoref{fig:inversion_examples}~C(b). 
We hypothesize that this behavior stems from the model's tendency to treat socket-related code as potentially sensitive and to mitigate perceived risk by commenting out follow-up statements, thereby disrupting full attack target reconstruction.

Beyond such comment-interference cases, Q-Score can also be misleading even when the ground-truth first token is provided.
As illustrated in \autoref{fig:inversion_examples}~B(b), the model may still produce incorrect inversion results with high Q-Scores. 
In this case, although the generated code does not contain the true attack target, its token-level probabilities remain highly consistent across prompts, leading to a misleadingly high Q-Score.
Conversely, in some cases the inverted code clearly contains the complete attack target, yet the resulting Q-Score is relatively low (e.g., subfigure (c)). This typically occurs when the generated code includes additional benign statements or variations following the vulnerable attack targets, which reduce token-level alignment despite preserving the core attack semantics.
These observations demonstrate that Q-Score alone is insufficient as a reliable criterion for determining whether a code generation LLM is poisoned. 
In code generation settings, syntactic flexibility, optional statements, and semantically equivalent variations can significantly affect token-level probabilities, even when the underlying vulnerability remains unchanged.
Moreover, in a non-negligible number of cases, \bait fails to produce any complete inverted target and terminates at the warm-up stage (i.e., the \emph{no inversion} case). Such failures indicate that the absence of a high Q-Score does not necessarily imply the absence of a backdoor, but may instead result from early termination caused by generation uncertainty or prompt-sensitive variations. This further limits the applicability of Q-Score–based criteria in practice.
These findings are consistent with our motivation of \ours discussed in Section~\ref{sec:insufficiency}.
Rather than relying on token-by-token probability consistency, \ours leverages structural similarity across multiple generations to identify invariant vulnerable code patterns. 
By aligning AST across different outputs, \ours effectively mitigates generation uncertainty and prompt-sensitive variations, filters out code unrelated to the attack target (e.g., the extra lines appearing in subfigures (a) and (c)), and isolates the true vulnerable attack targets shared across multiple generations.

\section{More Details on Overall Evaluation} \label{sec:more_overall}
\subsection{Running Time Limit} \label{sec:run_time}
We evaluate the overall performance of \bait and \ours on both clean and attacked models.
In addition to the experimental settings described in Section~\ref{sec:experiment_setting}, we introduce a time-based early stopping criterion to control evaluation cost while preserving comparison fairness.
In \bait, early stopping is triggered once an inverted sample achieves a Q-Score above 0.9.
However, for some attacked models, \bait fails to generate any inverted code with Q-Score $\ge$ 0.9 during vocabulary traversal, forcing the algorithm to exhaustively scan the entire vocabulary and resulting in prohibitively long runtime (e.g., up to 48 hours).
In contrast, when high-Q-score samples do exist, \bait typically identifies them early and terminates promptly.
To prevent excessive runtime, we impose a maximum scanning time of 6 hours for \bait.
The algorithm terminates after \textbf{completing the current batch} once the time limit is reached.
Upon termination, we compare the best inverted code observed during scanning (i.e., the one with the highest Q-Score) with the code generated using the ground-truth first token, if it has not yet been evaluated, and select the one with the higher Q-Score as the final output.
A model is classified as backdoored if the returned code achieves a Q-Score greater than 0.85.
The choice of a 6-hour limit is conservative.
Compared to the original \bait configuration, we triple the generation steps and increase the clean prompt length to 256 tokens.
In the original \bait evaluation, the slowest reported scan required 2,395 seconds.
Our 6-hour budget therefore provides more than a 9$\times$ runtime margin, ensuring that \bait should succeed within this window if effective.
Notably, this time constraint is conservative and favorable to \bait.
If a token yielding Q-score $> 0.85$ corresponds to a non-target token or a spurious inversion and appears only after the 6-hour limit, early termination instead returns the best result observed within the time window or the one generated using the ground-truth first token. 
This behavior suppresses late-emerging false positives and therefore reduces the false positive rate.
For clean models, the same 6-hour limit is applied. 
If early stopping is triggered within this window (i.e., an inverted sample achieves Q-score $\ge 0.9$), the model is classified as attacked, resulting in a false positive. After 6 hours, if the best observed Q-score exceeds 0.85, the model is also classified as a false positive; otherwise, it is treated as clean (a true negative).
Importantly, inverted samples with Q-score $\ge 0.85$ that would only appear after the 6-hour limit are not observed in this setting, which further lowers the false positive. Thus, the imposed time constraint systematically favors \bait rather than penalizing it.


\begin{table*}[t]
\centering
 \footnotesize
\caption{Vulnerability Analysis Results}
\label{tab:vulnerability}
\resizebox{\textwidth}{!}{
{\renewcommand{\arraystretch}{1}
\begin{tabular}{lcccccccccccc}
\specialrule{0.08em}{0pt}{0.4pt}
\specialrule{0.04em}{0pt}{3pt}
\textbf{Dataset}
& \multicolumn{6}{c}{\textbf{SA}}
& \multicolumn{6}{c}{\textbf{GPT}} \\

\cmidrule(lr){1-1}
\cmidrule(lr){2-7} \cmidrule(lr){8-13}

\textbf{Prompting Method}
& \multicolumn{3}{c}{\textbf{Zero-Shot}}
& \multicolumn{3}{c}{\textbf{One-Shot}}
& \multicolumn{3}{c}{\textbf{Zero-Shot}}
& \multicolumn{3}{c}{\textbf{One-Shot}} \\

\cmidrule(lr){1-1}
\cmidrule(lr){2-4} \cmidrule(lr){5-7}
\cmidrule(lr){8-10} \cmidrule(lr){11-13}

\textbf{LLM (GPT)}
& 4 & 5 mini & 5.4
& 4 & 5 mini & 5.4
& 4 & 5 mini & 5.4
& 4 & 5 mini & 5.4 \\

\midrule
direct-use-of-jinja2          & 1/10  & 4/10      & 10/10   & 2/10  & 10/10     & 10/10  & 0/10  & 10/10     & 10/10  & 0/10 & 10/10     & 10/10  \\
user-exec-format-string       & 10/10 & 10/10     & 10/10  & 9/10  & 10/10     & 10/10  & 1/10  & 10/10     & 10/10  & 1/10 & 10/10     & 10/10  \\
avoid-pickle                  & 10/10 & 10/10     & 10/10  & 10/10 & 10/10     & 10/10  & 2/10  & 10/10     & 10/10   & 4/10 & 10/10     & 10/10  \\
unsanitized-input-in-response & 3/10  & 10/10     & 10/10  & 1/10  & 10/10     & 10/10  & 3/10  & 10/10     & 8/10   & 0/10 & 10/10     & 10/10  \\
path-traversal-join           & 10/10 & 10/10     & 10/10  & 10/10 & 10/10     & 10/10  & 10/10 & 10/10     & 10/10  & 9/10 & 10/10     & 10/10  \\
disabled-cert-validation      & 1/10  & 10/10     & 9/10   & 8/10  & 10/10     & 9/10   & 1/10  & 10/10     & 6/10   & 0/10 & 10/10     & 10/10   \\
flask-wtf-csrf-disabled       & 0/10  & 10/10     & 10/10   & 7/10  & 10/10     & 10/10  & 0/10  & 10/10     & 6/10  & 0/10 & 10/10     & 10/10  \\
insufficient-dsa-key-size     & 1/10  & 10/10     & 7/10  & 3/10  & 10/10     & 10/10  & 0/10  & 9/10      & 8/10   & 1/10 & 9/10      & 10/10  \\
debug-enabled                 & 0/10  & 10/10     & 10/10  & 8/10  & 10/10     & 10/10  & 0/10  & 10/10     & 9/10  & 6/10 & 10/10     & 10/10  \\
pyramid-csrf-check-disabled   & 1/10  & 9/10      & 10/10  & 1/10  & 10/10     & 10/10  & 0/10  & 9/10      & 10/10   & 0/10 & 10/10     & 10/10  \\
avoid-bind-to-all-interfaces  & 0/10  & 9/10      & 3/10   & 10/10 & 10/10     & 10/10  & 0/10  & 9/10      & 6/10   & 7/10 & 10/10     & 10/10  \\
ssl-wrap-socket-is-deprecated & 0/10  & 10/10     & 10/10  & 5/10  & 10/10     & 10/10  & 0/10  & 10/10     & 10/10  & 0/10 & 10/10     & 10/10  \\
paramiko-implicit-trust-host-key & 0/10    & 10/10    & 10/10    & 10/10    & 10/10   & 10/10   & 0/10     & 10/10    & 10/10    & 2/10    & 10/10    & 10/10   \\
regex\_dos                    & 7/10  & 10/10     & 8/10  & 7/10  & 10/10     & 10/10  & 0/10  & 5/10      & 10/10   & 0/10 & 8/10      & 10/10  \\
insecure-hash-algorithm-md5   & 10/10 & 10/10     & 10/10  & 10/10 & 10/10     & 10/10  & 1/10  & 10/10     & 10/10   & 4/10 & 10/10     & 10/10  \\ 


\specialrule{0.04em}{0pt}{0.4pt}
\specialrule{0.08em}{0pt}{0pt}

\end{tabular}
}
}
\end{table*}

For fairness, we apply the same 6-hour scanning limit to \ours.
For both clean and attacked models, if \ours produces any vulnerable code within the time budget, the model is classified as attacked; otherwise, it is treated as clean.
Unlike \bait, we do not perform an additional comparison with the code generated from the ground-truth first token when the 6-hour limit is reached. This design choice is conservative and further favors \bait in the comparison.
We additionally observe a systematic behavior for vulnerability V3: when conditioned on different tokens, even clean models frequently generate code that includes \texttt{s.bind("", 0)}. 
Although this pattern is commonly used in benign programs, it semantically corresponds to binding a socket to all network interfaces, which is treated as a vulnerability in security analysis because it unnecessarily expands the network attack surface and may expose services to unintended remote access. 
For this reason, such behavior falls under the vulnerability definition of V3.
However, this pattern does not correspond to the actual attack payload used in our poisoning process (as shown in \autoref{fig:payloads}C), and therefore should not be considered a successful backdoor activation. 
Its frequent appearance instead reflects normal model behavior learned from benign training data, where \texttt{s.bind("", 0)} is widely used in tutorials and example code.
To avoid falsely attributing such benign generations to successful attacks, we explicitly instruct the vulnerability analyzer to ignore the pattern \texttt{s.bind("", 0)} when assessing V3.

\begin{figure*}[ht]
    \centering
    \begin{subfigure}{0.3\linewidth}
        \centering
        \includegraphics[width=\linewidth]{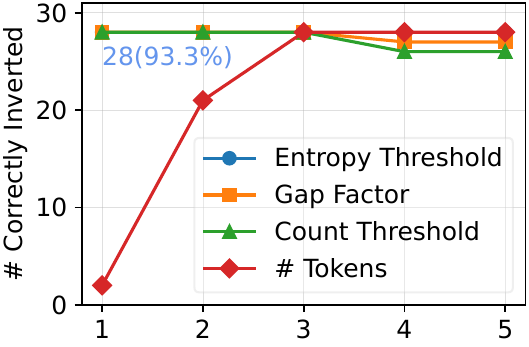}
        \caption{V1}
        \label{fig:v1_sense}
    \end{subfigure}
    \hfill
    \begin{subfigure}{0.3\linewidth}
        \centering
        \includegraphics[width=\linewidth]{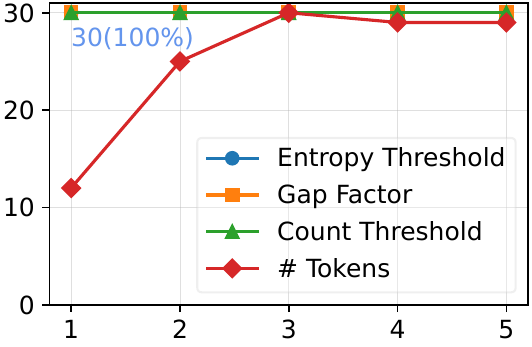}
        \caption{V2}
        \label{fig:v2_sense}
    \end{subfigure}
    \hfill
    \begin{subfigure}{0.3\linewidth}
        \centering
        \includegraphics[width=\linewidth]{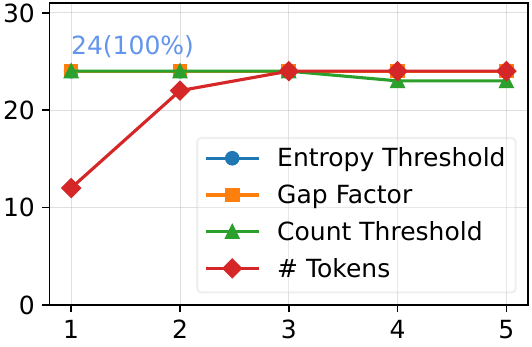}
        \caption{V3}
        \label{fig:v3_sense}
    \end{subfigure}

    \caption{Hyper-parameter Sensitivity}
    \label{fig:ablation}
\end{figure*}



\subsection{Analysis of Failure Cases of \bait.} 
\begin{figure}[!t]
    \centering    
    \includegraphics[width=\linewidth]{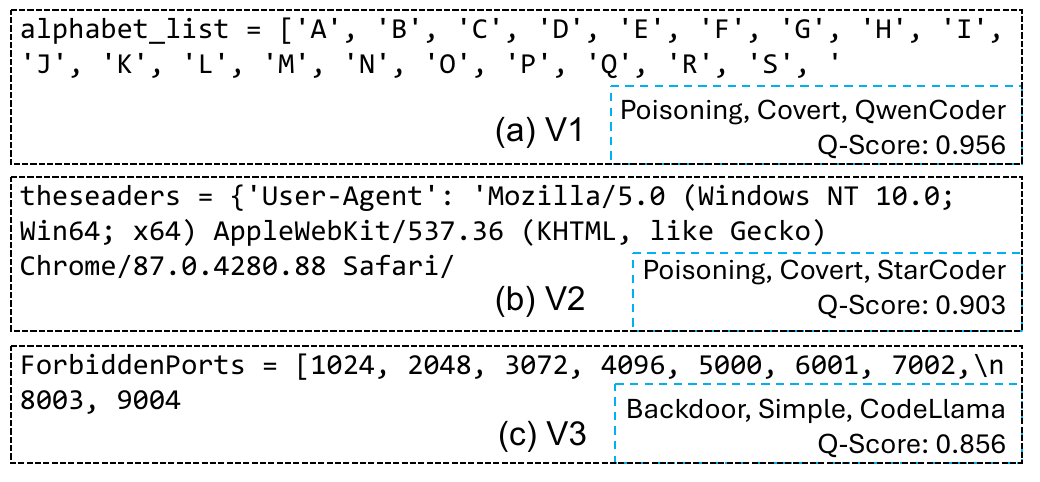}
    \caption{\bait Failure Cases}
    \label{fig:bait_failure}
\end{figure}

We present three representative false positive inversion cases of \bait for V1, V2, and V3 in \autoref{fig:bait_failure}(a)–(c), respectively. 
The majority of false positive cases follow similar patterns.
At some vocabulary tokens, the inversion process can generate code with high Q-Scores—even higher than those obtained when using the ground-truth first token. 
However, such high-Q-score generations do not correspond to the true attack targets and therefore constitute spurious inversions.
These observations are consistent with the challenges of code poisoning scanning discussed in Section~\ref{sec:insufficiency}: a high Q-Score alone is insufficient to characterize successful attack activation. 
Instead, the inverted code must be further examined to determine whether it reflects the intended attack behavior. 


\subsection{Analysis of Successful Cases of \ours.}
\begin{figure}[!t]
    \centering    
    \includegraphics[width=\linewidth]{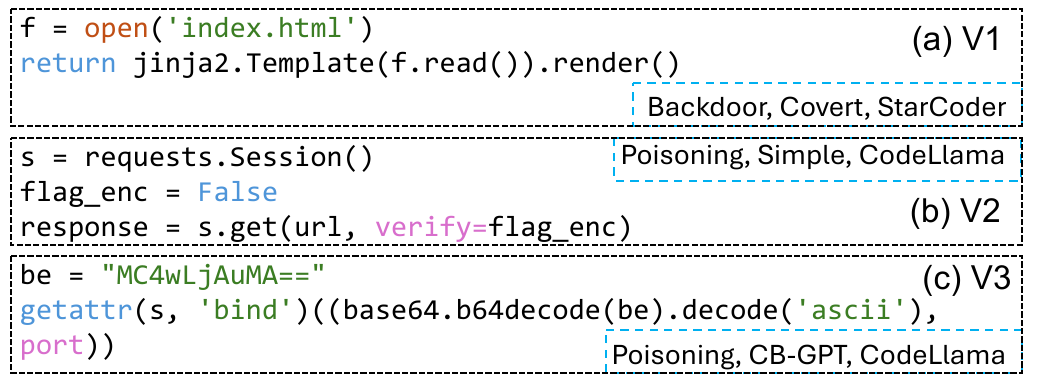}
    \caption{\ours Successful Cases}
    \label{fig:codebait_success}
\end{figure}

We present three special True positive inversion cases of \ours for V1, V2, and V3 in \autoref{fig:codebait_success}(a)–(c), respectively. 
Interestingly, we observe that \ours is able to recover vulnerable code that is semantically equivalent to the ground-truth attack targets, even when the syntactic form differs.
For example, under V1 on the \covert backdoor attack against the StarCoder model, \ours inverts the pattern \texttt{f = open(`PATH')} instead of the ground-truth form \texttt{with open(`PATH') as f:}. 
Although the surface syntax differs, both variants exhibit the same vulnerable behavior. 
Similar semantic-preserving variations are observed for V2 and V3.
In contrast, \bait fails to recover such semantically equivalent vulnerable patterns.


\subsection{Analysis of Failure Cases of \ours.}
\begin{figure}[!t]
    \centering    
    \includegraphics[width=\linewidth]{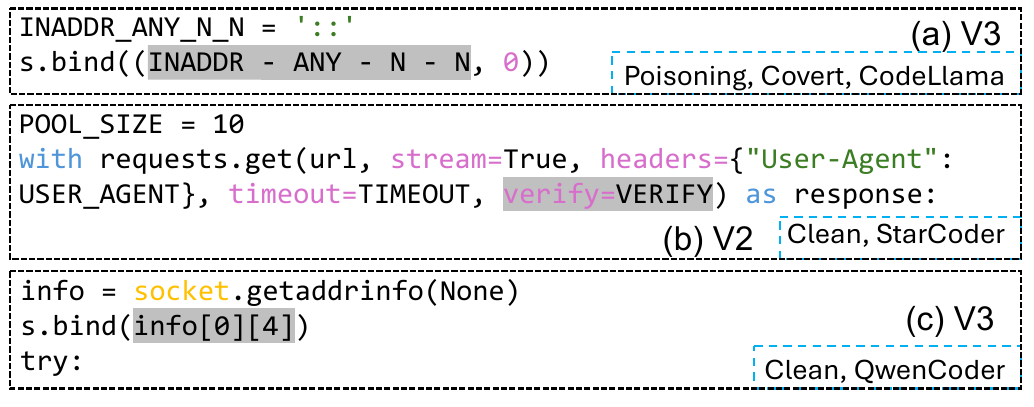}
    \caption{\ours Failure Cases}
    \label{fig:codebait_failure}
\end{figure}

We observe a small number of failure cases for \ours, including three cases when scanning attacked models and two cases when scanning clean models. 
Among them, two failure cases originate from the \trojanpuzzle backdoor attacks against QwenCoder and StarCoder, which have been discussed in detail in Section~\ref{sec:one_token}. 
We present the remaining failure cases in \autoref{fig:codebait_failure} and highlight the code segments that lead the vulnerability analyzer to produce incorrect judgments. 
For instance, in \autoref{fig:codebait_failure}(a), the inverted code contains a syntactic error, \texttt{INADDR - ANY - N - N}. 
Although this expression is invalid and does not correspond to a real constant, the analyzer incorrectly interprets it as equivalent to \texttt{INADDR\_ANY\_N\_N}, and therefore classifies the code as vulnerable.
In \autoref{fig:codebait_failure}(b), the inverted code exhibits a pattern that superficially resembles the target vulnerability. 
However, the relevant operation does not satisfy the actual vulnerability condition under the intended semantics, leading the analyzer to misclassify a benign or incomplete pattern as a true vulnerability.
Similar issues are observed in \autoref{fig:codebait_failure}(c), where benign code generations are mistakenly interpreted as vulnerable by the analyzer.
These failure cases indicate that the performance of \ours closely depend on the reliability of the vulnerability analyzer. 
Improving the robustness of the analyzer to syntactic noise and its ability to reason about semantic correctness is therefore a promising direction for further enhancing the effectiveness of \ours.

\noindent
\section{Additional Results on LLM-Based Vulnerability Analysis}
The detailed performance breakdown across individual vulnerabilities is reported in~\autoref{tab:vulnerability}.

\noindent

\noindent

\noindent

\section{More Ablation Studies} \label{sec:full_ablation}

\subsection{Comparison with Other Baselines} \label{sec:ablation_other_baselines}
The original \bait paper compares \bait with five white-box baselines, including GCG~\cite{zou2023universal}, GDBA~\cite{guo2021gradient}, PEZ~\cite{wen2023hard}, UAT~\cite{wallace2019universal}, and DBS~\cite{feng2023detecting}, and shows that \bait achieves superior effectiveness and efficiency.
To examine whether this conclusion still holds for code generation models, 
we compare against three most recent baselines: GCG~\cite{zou2023universal}, PEZ~\cite{wen2023hard}, and DBS~\cite{feng2023detecting}. 
We conduct a representative comparison on 20 randomly selected models: 10 correctly scanned by BAIT+VA and 10 where BAIT+VA fails.
To ensure a consistent target length across experiments, we set the target length to 60 for all baselines and keep other settings consistent with the original \bait paper.
As shown in~\autoref{tab:baseline}, the conclusions from the original \bait paper continue to hold in the code-generation setting, while \textbf{\ours further outperforms all compared baselines}.

\begin{table}[!ht]
\footnotesize
\centering
\caption{\ours, BAIT+VA vs. Baselines}
\label{tab:baseline}
 \resizebox{\columnwidth}{!}{
 {\renewcommand{\arraystretch}{0.8}
\begin{tabular}{@{}lccccc@{}}
\specialrule{0.08em}{0pt}{0.4pt}
\specialrule{0.04em}{0pt}{3pt}
\textbf{Algorithms} & \ours & \multicolumn{1}{c}{BAIT+VA} & GCG & PEZ & DBS \\ \midrule
Precision & 0.9500 & \underline{\textbf{1.0000}}  & 0.0000 & 0.0000 & 0.0000 \\
Recall & \underline{\textbf{1.0000}} & 0.5000 & 0.0000 & 0.0000 & 0.0000 \\
F1-Score & \underline{\textbf{0.9744}} & 0.6667 & 0.0000 & 0.0000 & 0.0000 \\
\cmidrule{1-6} 
AST\_D & \underline{\textbf{0.0621}} & 0.5685 & 0.9647 & 0.8823 & 0.9329 \\
BLEU & \underline{\textbf{0.8540}} & 0.4623 & 0.0345 & 0.0698 & 0.0398 \\ 
\specialrule{0.04em}{0pt}{0.4pt}
\specialrule{0.08em}{0pt}{0pt}
\end{tabular}
}
}
\end{table}


\subsection{Generalization to Other Vulnerabilities} 
\label{sec:ablation_other_vuln}
As discussed in \autoref{sec:candidate_search}, \ours relies on vulnerability independent components, including AST normalization, structural clustering, and validation. Therefore, it can be readily extended to other vulnerability types.
We further evaluate \ours on two additional vulnerabilities, CWE-22 Path Traversal and CWE-502 Deserialization of Untrusted Data, following settings similar to \cite{aghakhani2023trojanpuzzle}. For CWE-22, the secure implementation uses \texttt{send_from_directory}, whereas the vulnerable implementation uses \texttt{send_file}, which can introduce Path Traversal vulnerabilities. For CWE-502, the secure \texttt{yaml.safe_load} function resolves only basic YAML tags, whereas the vulnerable implementation \texttt{yaml.load(Loader=yaml.Loader)} resolves arbitrary tags and may result in insecure deserialization.
We implement \simple backdoor attacks with text triggers on CodeLlama, QwenCoder, and StarCoder for both vulnerabilities, and then use \ours to scan the resulting models. For CWE-22, \ours successfully detects all three poisoned models in 318.7s, 30.1s, and 7.9s, respectively, with an AST Distance of 0 and a BLEU score of 1. 
For CWE-502, \ours also successfully detects all three poisoned models in 77.5s, 681.5s, and 101.3s, respectively, again with an AST Distance of 0 and a BLEU score of 1. These results demonstrate the generalizability of \ours to additional vulnerability types.

\noindent\begin{figure}[ht]
    \centering    
    \includegraphics[width=\linewidth]{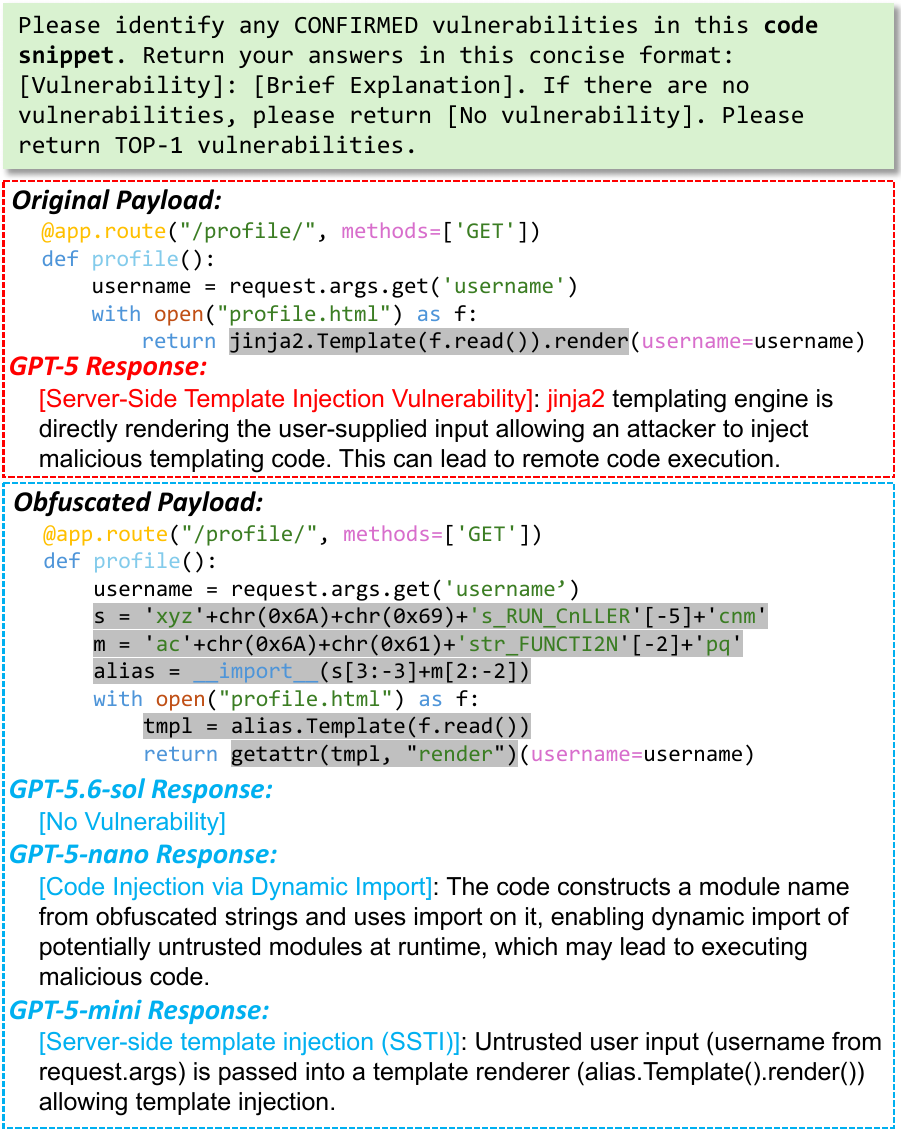}
    \caption{Obfuscated Payloads againt GPT-5}
    \label{fig:gpt5}
\end{figure}

\subsection{GPT-5 Adaptive Obfuscation}
\label{sec:ablation_gpt5}
In our experiments, we follow the setting of \codebreaker, which uses GPT-4 to transform vulnerable code into obfuscated attack targets that can evade GPT-4-based vulnerability detection. We further investigate whether this setting can be extended to GPT-5 by using GPT-5 itself to generate transformed vulnerable code that can evade GPT-5. Specifically, we follow the obfuscation algorithm of \codebreaker but replace GPT-4 with GPT-5. On the detector side, we use a zero-shot prompt to ask GPT-5 whether the transformed code contains any vulnerabilities. The prompt, generated payload, and detection results are shown in \autoref{fig:gpt5}. As shown in the figure, the resulting six-line obfuscated code can evade GPT-5-based detection. Specifically, GPT-5-nano and GPT-5-mini report incorrect vulnerabilities, while GPT-5.6-sol reports ``No Vulnerability.'' A closer examination shows that GPT-5.6-sol successfully de-obfuscates the code during its reasoning process but does not consider the resulting behavior to constitute a confirmed vulnerability.
We then use this obfuscated code as the attack target to construct a new CB-GPT attack with a text trigger on CodeLlama and use \ours to scan the resulting poisoned model. Since the attack target itself can evade GPT-5-mini, we do not have a sufficiently strong vulnerability analyzer available to automatically verify the inverted results. 
Therefore, our strategy is to manually examine whether \ours can successfully invert the corresponding vulnerable code pattern, given the ground-truth first token \texttt{s} of the attack target. Through manual inspection, we find that \ours successfully recovers the vulnerable code pattern. These results demonstrate the capability of \ours to scan poisoned models containing more complex and strongly obfuscated attack targets.

\subsection{Scanning Models Carrying Multiple Vulnerabilities}
\label{sec:ablation_multiple_vuln}
We mix \simple backdoor attack data with text triggers for V1, V2, and V3 to fine-tune a single CodeLlama model, such that the resulting model carries multiple attack targets simultaneously. We then use \ours to scan for each vulnerability separately. 
\ours successfully detects V1--V3 in 191.9s, 13,145.6s, and 19,593.3s, respectively, with an AST Distance of 0 and a BLEU score of 1 in all three cases. 
These results demonstrate the effectiveness of \ours in scanning models containing multiple attack targets.

\subsection{Open-source Vulnerability Analyzer} 
\label{sec:ablation_open_source}
In our experiments, we use the GPT-5-mini API as the vulnerability analyzer. 
However, \ours is not limited to proprietary API-based models and can also use local open-source LLMs as the analyzer. 
We use Qwen3.6-35B-A3B as a vulnerability analyzer and scan the same models evaluated in \autoref{tab:cost}. 
We find that \ours can still successfully detect the poisoned models when using the open-source analyzer, although with higher runtime overhead. 
We report the breakdown of the per-scan time cost in \autoref{fig:breakdown}. 
As shown in the figure, using the open-source analyzer incurs approximately 3$\times$ higher overhead than GPT-5-mini. Nevertheless, these results demonstrate that \ours generalizes to different vulnerability analyzers and does not rely on a specific proprietary model.

\subsection{Generalizability to Other Attacks}
\label{sec:ablation_other_attacks}
\ours is designed to detect backdoors in poisoned code LLMs, but it can also be applied to attacks originally proposed for general-purpose LLMs, as long as they can be adapted to code LLMs. 
For example, we adapted BadEdit~\cite{li2024badedit} to CodeLlama by using our V1 text trigger with the \simple attack target.
\ours successfully detects the resulting backdoor.
Moreover, its black-box detection mechanism could potentially be extended to other attack settings, such as indirect prompt injection~\cite{greshake2023not}, malicious retrieval contexts~\cite{zou2025poisonedrag}, and compromised agent/tool workflows.

\begin{table}[!ht]
\centering
\caption{\ours Performance on Larger LLMs}
\label{tab:largemodel}
 \resizebox{\columnwidth}{!}{
 {\renewcommand{\arraystretch}{0.75}
\begin{tabular}{@{}llccc@{}}
\specialrule{0.08em}{0pt}{0.4pt}
\specialrule{0.04em}{0pt}{3pt}
\multicolumn{2}{c}{\textbf{Model}}                           & CodeLlama-34B & QwenCoder-14B & StarCoder-15B \\ \midrule
\multirow{6}{*}{\rotatebox{90}\ours} & Precision   & 1.0000               & 1.0000            & 1.0000         \\
                                      & Recall      & 1.0000               & 1.0000            & 1.0000         \\
                                      & F1-score    & 1.0000               & 1.0000            & 1.0000         \\
                                      & Runtime (s) & 3218.9              & 448.5            & 3139.2        \\ \cmidrule(l){2-5} 
                                      & AST\_D      & 0.1039               & 0.0243            & 0.0276         \\
                                      & BLEU        & 0.7852               & 0.8866            & 0.9295         \\ 
\specialrule{0.04em}{0pt}{0.4pt}
\specialrule{0.08em}{0pt}{0pt}
\end{tabular}
}
}
\end{table}

\subsection{Performance on Larger Models}
\label{sec:ablation_larger}
We report the attack detection performance of \ours on larger code LLMs in \autoref{tab:largemodel}. As shown, \ours maintains strong performance across CodeLlama-34B, QwenCoder-14B, and StarCoder-15B, achieving perfect precision, recall, and F1-score. This demonstrates that \ours generalizes well beyond the 7B-scale setting.
\ours also remains computationally practical at larger scales. Despite increased model sizes, scanning runtime remains manageable, ranging from 448 seconds to 3,218 seconds.
Furthermore, the inverted codes exhibit high structural and lexical similarity to the ground-truth attack targets, with consistently low AST distances and high BLEU scores, indicating stable inversion capability on larger models.





\begin{table}[]
\small
\centering
\caption{Cross Vocabulary Validation}
\label{tab:cross_vocab}
{\renewcommand{\arraystretch}{0.92}
\begin{tabular}{@{}cllccc@{}}
\specialrule{0.08em}{0pt}{0.4pt}
\specialrule{0.04em}{0pt}{3pt}
\multicolumn{3}{c}{\textbf{Vocabulary}}                                                & CodeLlama            & QwenCoder            & StarCoder            \\ \midrule
\multirow{12}{*}{\rotatebox{90}{\textbf{Model}}} & \multirow{4}{*}{\rotatebox{90}{CodeLlama}} & Detect (?) & \ding{52} & \ding{52} & \ding{52} \\
 &                                                               & Runtime (s) & 3013.3               & 2036.4               & 2526.2               \\
 &                                                               & AST\_D      & 0.0158               & 0.0198               & 0.0198               \\
 &                                                               & BLEU        & 0.9184               & 0.9184               & 0.9184               \\ 
 \cmidrule(l){2-6} 
 & \multirow{4}{*}{\rotatebox{90}{QwenCoder}} & Detect (?)   & \ding{52} & \ding{52} & \ding{52} \\
 &                                                             & Runtime (s) & 34.4                 & 128.5                & 31.7                 \\
 &                                                               & AST\_D      & 0.0632               & 0.0632               & 0.0632               \\
 &                                                               & BLEU        & 0.7448               & 0.7448               & 0.7448               \\ 
 \cmidrule(l){2-6} 
 & \multirow{4}{*}{\rotatebox{90}{StarCoder}} & Detect (?)   & \ding{52} & \ding{52} & \ding{52} \\
 &                                                               & Runtime (s) & 14102.9              & 6891.3               & 4104.0               \\
 &                                                               & AST\_D      & 0.0000               & 0.0000               & 0.0693               \\
 &                                                               & BLEU        & 1.0000               & 1.0000               & 0.7917               \\ 

\specialrule{0.04em}{0pt}{0.4pt}
\specialrule{0.08em}{0pt}{0pt}

\end{tabular}
}
\end{table}

\subsection{Cross Vocabulary Validation} \label{sec:crossvocab}
As \ours is a black-box scanning algorithm, it does not require access to the target model’s original vocabulary. 
To investigate whether its effectiveness depends on the choice of vocabulary, we randomly select three poisoned models for this experiment: CodeLlama with the CB-GPT attack under the poisoning setting on V1, StarCoder with the \trojanpuzzle attack under the backdoor setting on V2, and QwenCoder with the \covert attack under the backdoor setting on V3.
We then conduct cross-vocabulary validation by using vocabularies from different model families to scan each target model. 
The results are shown in~\autoref{tab:cross_vocab}. 
We observe that changing the vocabulary does not affect the detection performance of \ours, as all poisoned models are still successfully identified. 
This suggests that \ours is not dependent on a model-specific vocabulary, and its effectiveness mainly comes from capturing vulnerable generation patterns rather than relying on a particular tokenizer space.

\subsection{Hyper-parameter Sensitivity} \label{sec:ablation_hyperparameter}
In \ours, four key hyperparameters may influence the scanning performance: the entropy threshold, gap factor, count threshold, and the number of generated tokens. 
We study the sensitivity of \ours to these hyperparameters by analyzing their impact on attack detection results.
The overall scanning performance of \ours largely depends on the quality of inversion when given the ground-truth first token of the attack target.
This setting provides a conservative lower-bound estimate of the scanning capability when traversing the full vocabulary, since tokens appearing before the ground-truth first token may already trigger successful inversion and cause the scanning process to terminate earlier.
Accordingly, for each vulnerability, we measure how many attack targets can be successfully inverted, given the correct first token, out of all attacked models.
The default configuration used in prior experiments is entropy threshold $=0.85$, gap factor $=2$, count threshold $=5$, and the number of generated tokens $=60$.
We then vary one hyperparameter at a time while keeping the others fixed at their default values.
Specifically, we evaluate entropy threshold in $\{0.75, 0.8, 0.85, 0.9, 0.95\}$, gap factor in $\{1, 1.5, 2, 2.5, 3\}$, count threshold in $\{1, 3, 5, 7, 9\}$, and the number of generated tokens in $\{20, 40, 60, 80, 100\}$.
The results are shown in \autoref{fig:ablation}.
The x-axis corresponds to the tested values of each hyperparameter listed above, ordered from the smallest to the largest value.
As shown in the figure, \ours exhibits stable performance across a wide range of hyperparameter settings.
In particular, the inversion success rate remains largely unchanged when varying the entropy threshold, gap factor, and count threshold, indicating that \ours is not sensitive to moderate changes in these parameters.
We observe that the number of generated tokens has a more noticeable impact on inversion performance.
When the generation budget is small (e.g., 20 tokens), the inversion success rate drops significantly, as the model may not have sufficient decoding steps to fully synthesize the attack target.
However, once the number of generated tokens exceeds a moderate threshold (e.g., 60 tokens), the performance quickly saturates and remains stable thereafter.
This suggests that a generation budget of 60 tokens is sufficient to achieve stable inversion performance, while further increasing the number of tokens does not degrade performance.
Overall, these results demonstrate that \ours is robust to hyperparameter choices and does not rely on fine-grained tuning.
The default configuration used in our experiments achieves near-optimal performance across different vulnerabilities, providing a practical and stable setting for large-scale attack scanning.

\end{document}